\documentclass[prd,aps,twocolumn,floats,superscriptaddress,nofootinbib,floatfix,notitlepage]{revtex4-1}
\usepackage{exscale}                  
\usepackage[intlimits]{amsmath}       
\usepackage{amsfonts}
\usepackage{amssymb,amscd}
\usepackage{wrapfig}
\usepackage{graphicx}
\usepackage{array}
\usepackage{bbm}
\usepackage{multirow}
\usepackage[T1]{fontenc}
\usepackage{times}
\usepackage[caption=false]{subfig}
\usepackage[x11names]{xcolor}
\usepackage{epstopdf}
\usepackage{slashed}
\usepackage{braket}
\usepackage{pstricks}
\usepackage{ulem}
\usepackage{comment}

\newcommand{\sn}{\, \mathrm{sn}}
\newcommand{\cn}{\, \mathrm{cn}}
\newcommand{\dn}{\, \mathrm{dn}}

\def\bx{{\mathbf x}}

\def\bk{{\mathbf k}}

\def\bp{{\mathbf p}}
\def\bE{{\mathbf E}}
\def\bB{{\mathbf B}}
\def\pr{\prime}
\def\cd{\cdot}

\begin{document}

\title{Transient anomalous charge production in strong-field QCD}

\author{Naoto Tanji}\email{tanji@thphys.uni-heidelberg.de}
\affiliation{Institut f\"{u}r Theoretische Physik, Universit\"{a}t
  Heidelberg, 69120 Heidelberg, Germany}
  
\author{Niklas Mueller}\email{n.mueller@thphys.uni-heidelberg.de}
\affiliation{Institut f\"{u}r Theoretische Physik, Universit\"{a}t
  Heidelberg, 69120 Heidelberg, Germany}

\author{J\"urgen Berges}\email{j.berges@thphys.uni-heidelberg.de}
\affiliation{Institut f\"{u}r Theoretische Physik, Universit\"{a}t
  Heidelberg, 69120 Heidelberg, Germany}

\begin{abstract}
We investigate axial charge production in two-color QCD out of equilibrium. We compute the real-time evolution starting with spatially homogeneous strong gauge fields, while the fermions are in vacuum. The idealized class of initial conditions is motivated by glasma flux tubes in the context of heavy-ion collisions.  We focus on axial charge production at early times, where important aspects of the anomalous dynamics can be derived analytically. This is compared to real-time lattice simulations. Quark production at early times leading to anomalous charge generation is investigated using Wilson fermions. Our results indicate that coherent gauge fields can transiently produce significant amounts of axial charge density, while part of the induced charges persist to be present even well beyond characteristic decoherence times. The comparisons to analytic results
provide stringent tests of real-time representations of the axial anomaly on the lattice.
\end{abstract}

\maketitle


\section{Introduction and overview\label{sec:introduction}}

Anomalous quantum processes violating classical symmetries play a crucial role for our understanding of fundamental properties of matter. A most prominent example concerns the question about the origin of the matter-antimatter asymmetry in the universe, which has long been discussed in terms of sphaleron baryogenesis, where the sphalerons denote the lowest-barrier configurations separating energy-degenerate minima in the electroweak theory~\cite{Kuzmin:1985mm,Arnold:1987mh,Dine:2003ax}. Similar configurations are expected to play an important role during the early stages of collision experiments with heavy nuclei, and may lead to an anomalous generation of electric currents from the so-called chiral magnetic effect in the context of the theory of the strong interaction~\cite{Kharzeev:2007jp,Fukushima:2008xe,Fukushima:2010vw,Kharzeev:2015znc}. The departure from thermal equilibrium is an essential ingredient in all scenarios of baryogenesis from
microphysical laws~\cite{Sakharov:1967dj} and is crucial for our understanding of the initial stages of very energetic heavy-ion collisions~\cite{Lappi:2006fp,Gelis:2010nm}.

For non-Abelian gauge theories in thermal equilibrium, 
sphaleron transitions are expected to dominate the late-time behavior of the Chern--Simons number 
associated to transitions between topologically distinct ground states. 
The non-perturbative computation of the thermal transition rate in the presence of these 
spatially-localized classical field configurations can be achieved using classical real-time 
simulation techniques on a lattice~\cite{Arnold:1996dy,Moore:1997cr}. 

Away from thermal equilibrium, the anomalous processes are in general
not related to thermal transitions between different vacua, but complicated out-of-equilibrium processes. Recently, this has been investigated in the context of early-universe electroweak baryogenesis from fast quench dynamics~\cite{Saffin:2011kn}, or highly populated gluon
fields characteristic for the initial stages of relativistic heavy-ion collisions~\cite{Mace:2016svc}. While these lattice studies concentrate on the behavior of the Chern--Simons number, out-of-equilibrium conditions can have dramatic consequences for the presence of transient anomalous effects that are not associated to transitions between topologically distinct ground states. 
A recent example concerns the phenomenon of anomaly-induced dynamical refringence in strong-field quantum electrodynamics despite its trivial vacuum structure~\cite{Mueller:2016}.

In this work, we investigate transient anomalous effects in quantum chromodynamics (QCD) with two colors. The aim is to gain \mbox{(semi-)analytical} insights into axial charge generation due to the Adler--Bell--Jackiw anomaly equation~\cite{Adler:1969gk,Bell:1969ts} 
\begin{align}
\partial_\mu j_5^\mu 
 = 2m\overline{\psi} i\gamma_5 \psi 
    +\frac{g^2}{4\pi^2} \mathbf{E}^a\cdot\mathbf{B}^a \label{eq:ABJ}
\end{align}
out of equilibrium. It relates the four-divergence of the axial current $j_5^\mu = \overline{\psi}\gamma^\mu \gamma_5 \psi$ ($\mu = 0,1,2,3$ with Dirac matrices  $\gamma^\mu$ and $\gamma_5 = i \gamma^0 \gamma^1 \gamma^2 \gamma^3$) to the mixing of the different chiral components of the fermion fields $\psi$ of mass $m$, and to the anomaly term $\sim \mathbf{E}^a\cdot\mathbf{B}^a$ involving the color electric fields $\mathbf{E}^a$ and magnetic fields $\mathbf{B}^a$ ($a=1,2,3$). 

To this end, we consider the real-time evolution starting with spatially homogeneous gauge fields. The field configuration is motivated by the glasma flux-tube scenario, where the gluonic gauge fields in the immediate aftermath of a heavy-ion collision are dominated by coherent longitudinal color-electric and magnetic fields~\cite{Lappi:2006fp}. For a sufficiently energetic collision, the relevant gauge coupling $g$ is weak and we consider $g\ll 1$. 

More precisely, we investigate an idealized class of initial conditions, where the expectation values at time $t=0$ for color-electric fields $E^a_i(0)$ and magnetic fields $B^a_i(0)$ with spatial components $i = x,y,z$ in temporal gauge are given by  
\begin{align}
\langle E_x^1(0) \rangle &=\langle E_y^2(0) \rangle = \langle E_z^3(0) \rangle \sim \frac{Q^2}{g}, \notag \\
\langle B_x^1(0) \rangle &= \langle B_y^2(0) \rangle = \langle B_z^3(0) \rangle =0, \label{eq:initcond}
\end{align} 
corresponding to an energy density $\sim Q^4/g^2$ parametrized in terms of the characteristic scale $Q$. All other modes, as well as the fermion sector, are taken to be in (free) vacuum initially.
While we start with zero macroscopic color-magnetic field such that the anomalous contribution vanishes initially, it is generated during subsequent times. The nonequilibrium classical time evolution of the Yang--Mills fields can be solved analytically~\cite{baseyan1979nonlinear,matinyan1981classical,biro1995chaos}. By taking into account quantum fluctuations, one observes that the solution represents the leading contribution for the corresponding quantum dynamics on a time scale shorter than $t_\Theta \sim Q^{-1} \ln g^{-2}$. This allows us to derive a closed-form expression for the early-time behavior of axial charge generation from the anomaly equation. 

The times beyond $t_\Theta$, after which fluctuations cause decoherence of the initially uniform fields, are no longer described by our analytical treatment linearizing in fluctuations. Using real-time lattice simulation techniques, we verify that the early-time lattice dynamics indeed accurately reproduces our analytical results. 
Furthermore we find that transient homogeneous fields can lead to nonzero axial charge density even beyond the characteristic decoherence time.  

While in general the investigation of more realistic field configurations and later times cannot be based on analytic solutions and requires non-perturbative real-time lattice simulation techniques~\cite{Aarts:1998td,Borsanyi:2008eu,Berges:2010zv,Saffin:2011kc,Saffin:2011kn,Hebenstreit:2013qxa,Kasper:2014uaa,Tanji:2015ata,Gelis:2015kya,Gelfand:2016prm}, 
our analytical expressions provide a stringent precision test for the numerical approaches. On a lattice, the axial anomaly is deeply connected to the fermion doubling problem and its regularization, which is well understood in Euclidean or `imaginary-time' lattice field theory~\cite{Karsten:1980wd,Nielsen:1980rz,Friedan:1982nk}. 
In particular, for Euclidean Wilson fermions all doublers can be regularized using a spatiotemporal Wilson term. In contrast, real-time simulations typically employ a combination of a spatial Wilson term together with a suppression of possible temporal doublers using suitable initial conditions~\cite{Saffin:2011kn,Saffin:2011kc}. Employing real-time lattice simulations for two-color QCD, we analyze in detail the validity of the axial anomaly equation on the lattice by explicitly computing the nonequilibrium axial charge density from the underlying fermion current.   

The paper is organized as follows. In section~\ref{sec:gauge}, we investigate the real-time evolution of the gauge field sector. We derive an analytic expression for the production of the axial charge and compare it to real-time lattice simulations in pure gauge theory. In section \ref{sec:fermion}, we investigate the fermion sector and perform real-time lattice simulations with Wilson fermions. We analyze the axial anomaly out of equilibrium and demonstrate that it can be accurately computed using a spatial Wilson term. Section \ref{sec:conclusion} is devoted to concluding remarks.
In an appendix, we show an alternative verification of the chiral anomaly with a cutoff regularization method.

\section{Transient anomalous charge production: gauge sector \label{sec:gauge}}

\subsection{Analytic discussion}

We consider a non-Abelian gauge theory with $SU(2)$ color gauge group.
Taking two colors simplifies the analysis as compared to the $SU(3)$ gauge group of
QCD, while for our aims the difference is of minor relevance. We do not consider longitudinally expanding systems, such as addressed in Ref.~\cite{Kharzeev:2001ev}. We concentrate on the nonequilibrium dynamics, starting from an initial state characterized by macroscopic color-electric fields of order $\langle E(0) \rangle \sim Q^2/g$ in the weak gauge coupling $g \ll 1$ relevant at a sufficiently high energy scale $Q$. All other gauge field modes, as well as the fermion sector, are taken to be in (free) vacuum initially. The early-time behavior for this problem can be solved analytically in an expansion in powers of the gauge coupling $g$ following along the lines of Ref.~\cite{Berges:2011sb}, where simpler initial conditions have been considered in the absence of anomalous corrections. In particular, at leading order in $g$ there is no back-reaction of the fermion sector on the gauge field dynamics at early times (see e.g.~\cite{Kasper:2014uaa}). 

It is convenient to formulate the gauge field dynamics in terms of gauge potentials $A^a_\mu(x)$ with $x = (x^0, {\mathbf x})$ in temporal gauge, where $A^a_0(x) = 0$, and to split the field into a time-dependent expectation value $\langle A^a_i(x) \rangle = \bar{A}^a_i(x^0)/g$ and a quantum fluctuation according to
\begin{equation}
A^a_i(x) \, = \, g ^{-1} \bar{A}^a_i(x^0) + \delta A^a_i(x) \, . 
\label{eq:rescaledA}
\end{equation}  
Introducing the rescaled macroscopic field $\bar{A}$ simplifies the power-counting in $g$. 
Starting from the spatially homogeneous macroscopic field configuration, we may linearize the dynamics in $\delta A$ for sufficiently early times. The range of times, for which the linearized description is valid, is determined below.    

At zeroth order in the fluctuations, we obtain the field equation for the macroscopic field 
\begin{equation}
\left(D_\mu[\bar{A}] F^{\mu\nu}[\bar{A}]\right)^a \, = \, 0 \, ,
\label{eq:backgroundfieldequation}
\end{equation}
which corresponds to the classical Yang--Mills equation with field strength tensor
\begin{equation}
F^a_{\mu\nu}[\bar{A}] \, = \, \partial_\mu \bar{A}^a_\nu - \partial_\nu \bar{A}^a_\mu -\epsilon^{abc} \bar{A}^b_\mu \bar{A}^c_\nu
\label{eq:fieldstrength}
\end{equation}
and covariant derivative
\begin{equation}
D_\mu^{ab}[\bar{A}] \, = \, \partial_\mu \delta^{ab} -\epsilon^{acb} \bar{A}_\mu^c \, .
\end{equation}
One observes that the classical equation (\ref{eq:backgroundfieldequation}) for the rescaled
macroscopic field $\bar{A}$ does not depend on the coupling $g$. Moreover, all spatial
derivatives of $\bar{A}$ actually vanish. The next order corresponds to the linearized equation for the fluctuations~\cite{Tudron:1980gq}, 
\begin{eqnarray}
&&\left(D_\mu[\bar{A}]D^\mu[\bar{A}]\delta A^\nu\right)^a - \left(D_\mu[\bar{A}]D^\nu[\bar{A}]\delta A^\mu\right)^a  \nonumber\\
&& -\epsilon^{abc} \delta A^b_\mu F^{c\mu\nu}[\bar{A}] \, = \, 0 \, .
\label{eq:fluctuationequation}
\end{eqnarray}
Equations (\ref{eq:backgroundfieldequation}) and (\ref{eq:fluctuationequation}) describe the gauge dynamics up to corrections of order $(\delta A)^2$ in the fluctuations and to leading order in the coupling $g$.

Writing $t \equiv x^0$, we consider the time-dependent field configuration~\cite{baseyan1979nonlinear,matinyan1981classical} 
\begin{align}
\bar{A}_i^a(t)=\mathcal{A}(t)\left(\delta^{a1}\delta_{ix}+\delta^{a2}\delta_{iy}+\delta^{a3}\delta_{iz} \right)  . \label{eq:fieldconf}
\end{align} 
The corresponding chromo-electric and magnetic field components are 
\begin{align}
\langle E_x^1 \rangle(t)=\langle E_y^2 \rangle(t)=\langle E_z^3 \rangle(t)=g^{-1}\partial_t\mathcal{A}(t) \, , \\
\langle B_x^1 \rangle(t)=\langle B_y^2 \rangle(t)=\langle B_z^3 \rangle(t)=g^{-1} \mathcal{A}^2(t) \, , 
\end{align}
from which we recover the initial conditions (\ref{eq:initcond}) by choosing
\begin{align}
\mathcal{A}(0) = 0 \, , \qquad \partial_t\mathcal{A}(0) = \frac{Q^2}{\sqrt{3}} \, .
\label{eq:AIC}
\end{align}  
With the employed normalization the energy density is given by $Q^4/2g^2$.

\begin{figure}[t]
\begin{center}
\includegraphics[width=8.0cm]{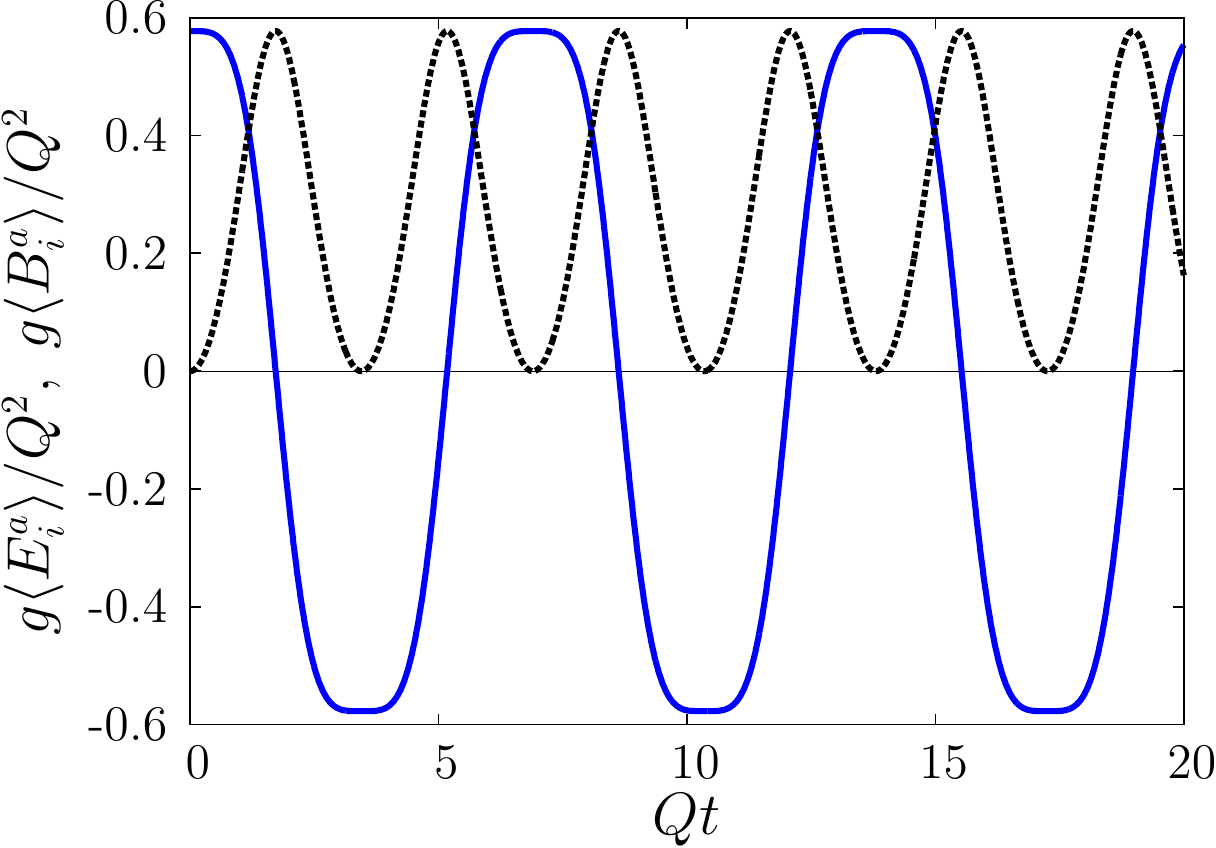}
\caption{Evolution of the rescaled nonzero components of the color-electric field (solid line)
 and magnetic field (dashed line) in the linear regime, where they are described by the analytic solutions (\ref{eq:ele}) and (\ref{eq:mag}). }
\label{fig:EandB}
\end{center}
\end{figure}

For the configuration (\ref{eq:fieldconf}), the macroscopic field equation (\ref{eq:backgroundfieldequation}) reads 
\begin{align}
\partial_t^2\mathcal{A}(t)+2 \mathcal{A}^3(t)=0 \, .\label{eq:homogenousYMEquation}
\end{align}
With the initial conditions (\ref{eq:AIC}), the solutions of this equation can be expressed in terms of Jacobi elliptic functions as
\begin{align}
\mathcal{A}(t)=\frac{Q}{3^{1/4}}\, \text{cn}\left(\sqrt{\frac{2}{\sqrt{3}}}\, Q t- K(1/2),\;\frac{1}{2}\right),\label{eq:solution_zerothorder}
\end{align}
where $K(1/2)$ denotes the complete elliptic integral of the first kind~\cite{abramowitz1965handbook}. 
The nonzero components of the color-electromagnetic fields then read
\begin{align}
\langle E_i^a  \rangle(t) \notag 
 = -&\sqrt{\frac{2}{3}} \frac{Q^2}{g} \sn \left(\sqrt{\frac{2}{\sqrt{3}}}\, Q t-K(1/2) ,\frac{1}{2} \right)\\\times
   &\dn \left(\sqrt{\frac{2}{\sqrt{3}}}\, Q t-K(1/2) ,\frac{1}{2} \right)  , \label{eq:ele} \\
\langle B_i^a  \rangle(t)
  = \frac{1}{\sqrt{3}}& \frac{Q^2}{g} \cn^2 \left(\sqrt{\frac{2}{\sqrt{3}}}\, Q t-K(1/2) ,\frac{1}{2} \right) . \label{eq:mag}
\end{align}
In Fig.~\ref{fig:EandB}, these fields are plotted as a function of time. They oscillate in time with a characteristic frequency $\sim Q$. By multiplying with $g/Q^2$, the quantities become dimensionless and independent of the value of $g$.  

Starting from the configuration with a strong color-electric field and vanishing magnetic field, one observes that the latter is subsequently generated. The build-up of the chromo-magnetic fields is possible because of the non-linear gauge dynamics, which is uniquely due to the non-Abelian nature of the theory. In general, one can write
\begin{align}
\mathbf{E}^a\cdot\mathbf{B}^a = - \epsilon^{\mu\nu\rho\sigma}\partial_\mu\;\text{tr}\left(A_\nu\partial_\rho A_\sigma+\frac{2}{3}igA_\nu A_\rho A_\sigma\right).\label{eq:anomaly_Terms}
\end{align}
Therefore, a nonzero $\mathbf{E}^a\cdot\mathbf{B}^a$ may be obtained even for spatially homogeneous gauge potentials in a non-Abelian theory. 

\begin{figure}[t]
\begin{center}
\includegraphics[width=8.0cm]{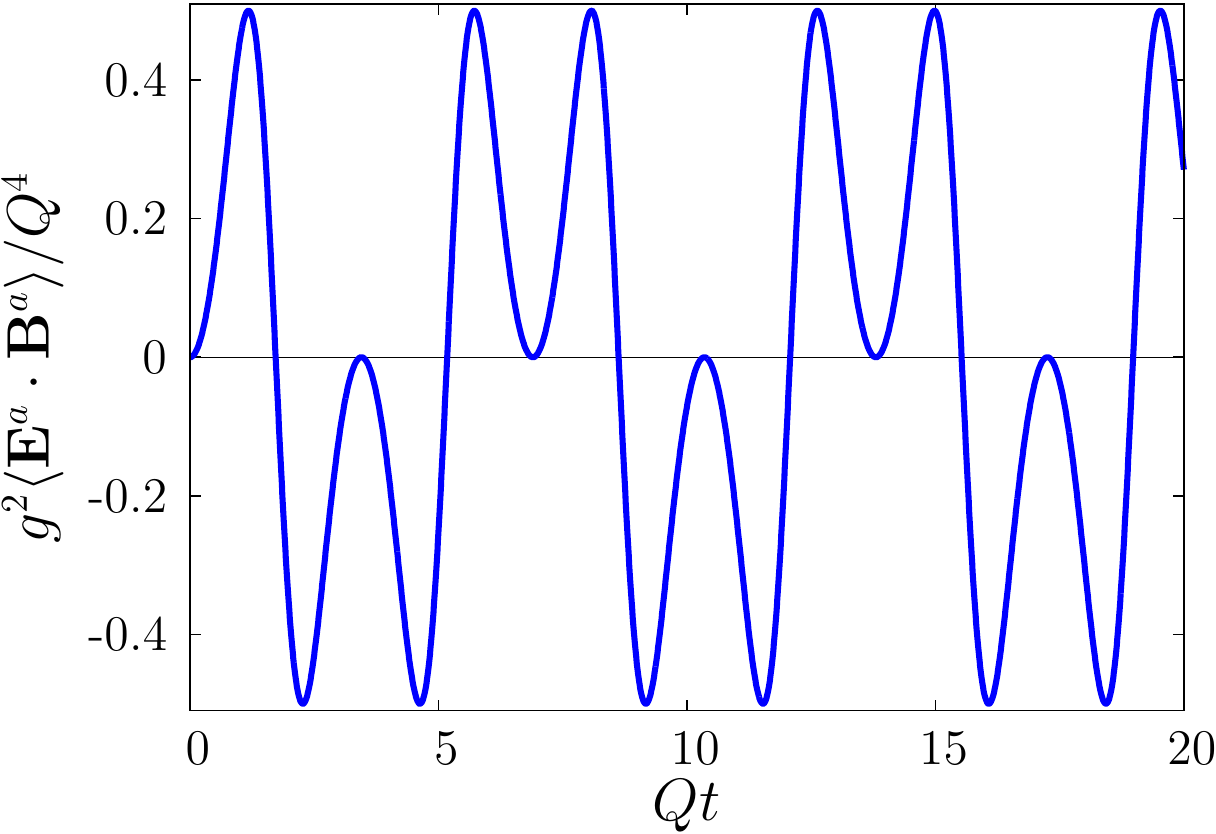}
\caption{Time dependence of the anomalous contribution $\sim \langle \mathbf{E}^a \cdot \mathbf{B}^a \rangle$ in the linear regime.  }
\label{fig:EdotB}
\end{center}
\end{figure}

\begin{figure}[t]
\begin{center}
\includegraphics[width=8.0cm]{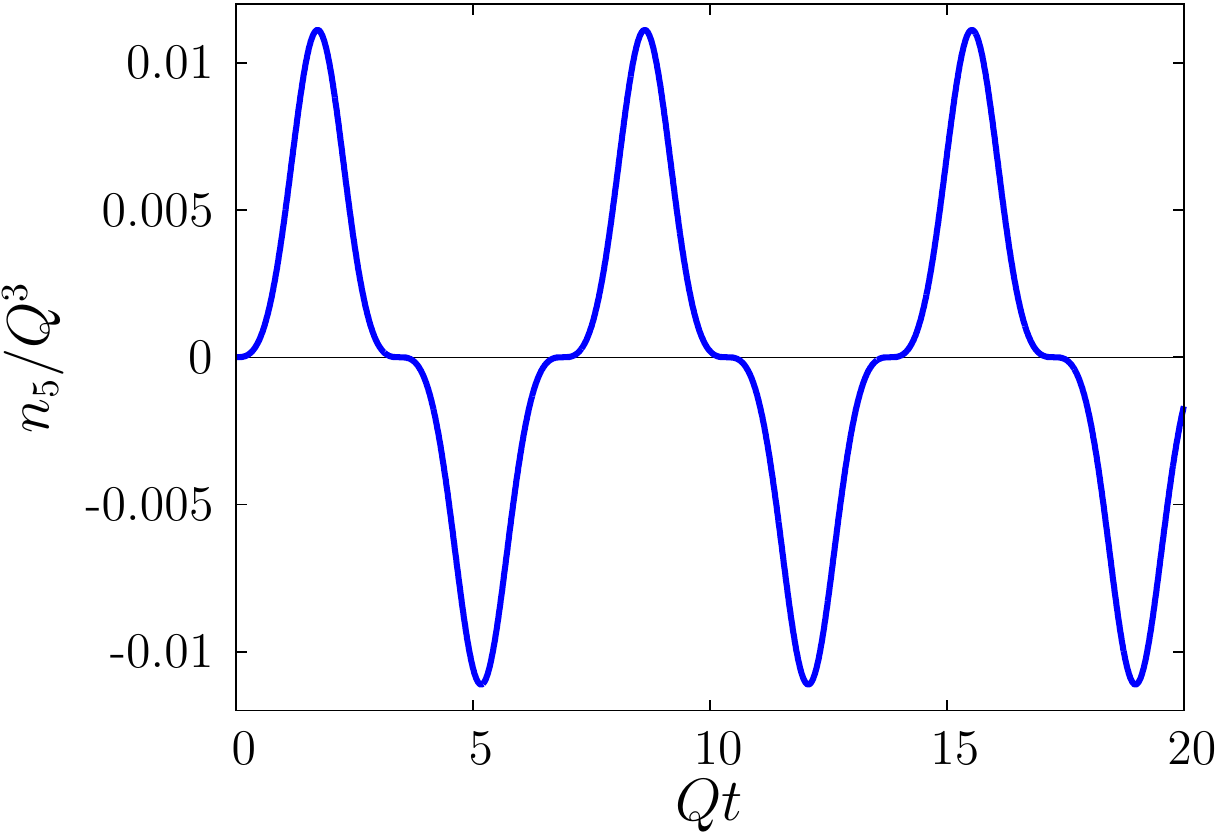}
\caption{Time dependence of the axial charge density in the massless limit, as described by the analytic result (\ref{eq:resultn5}).}
\label{fig:analytic_n5}
\end{center}
\end{figure}

The anomaly equation (\ref{eq:ABJ}) relates $\mathbf{E}^a \cdot \mathbf{B}^a$ to the four-divergence of the axial fermion current $j^\mu_5$. For the homogeneous system, the spatial divergence drops out for the evaluation of the expectation value of this current. The axial charge density $n_5 (t)=\langle j_5^0 (x) \rangle$ is then obtained by integrating over time:
\begin{equation} \label{eq:intABJ}
n_5 (t) = 2m \int_0^t dt^\pr \, \langle \overline{\psi} i\gamma_5 \psi \rangle  (t^\pr )
 +\frac{g^2}{4\pi^2} \int_0^t dt^\pr \, \langle\mathbf{E}^a \cdot \mathbf{B}^a \rangle (t^\pr )\,  
\end{equation}
for zero initial axial charge. The first term on the right hand side arises from the mixing of the different chiral field components in the presence of a mass. Therefore, in a massless theory the axial charge density is entirely determined by the anomalous second term. In particular,
to lowest order in the fluctuations we have $\langle \mathbf{E}^a \cdot \mathbf{B}^a\rangle =
\langle \mathbf{E}^a \rangle \cdot \langle \mathbf{B}^a\rangle$, and the time evolution of this term is plotted in Fig.~\ref{fig:EdotB}. Therefore, in this approximation our solutions (\ref{eq:ele}) and (\ref{eq:mag}) determine the dynamics of the axial charge generation, and in the massless limit we find from integration:
\begin{equation}
n_5 (t) = \frac{Q^3}{3^{3/4} 4\pi^2} \cn^3 \left(\sqrt{\frac{2}{\sqrt{3}}}\, Q t -K(1/2),\frac{1}{2} \right) \, , 
\label{eq:resultn5}
\end{equation}
as plotted in Fig.~\ref{fig:analytic_n5}. We further find that only the second term $\sim A^3$ on the right hand side of (\ref{eq:anomaly_Terms}) 
contributes to the anomaly for the initial conditions considered, such that (\ref{eq:resultn5})
can also be written as
\begin{align}
n_5(t)=\frac{1}{4\pi^2} \mathcal{A}^3(t) \, .
\end{align}

Equation (\ref{eq:resultn5}) is the central result of this section, which will be further discussed and used in section~\ref{sec:fermion} to verify implementations of the axial anomaly in real-time lattice simulations. However, before doing so we have to establish the solution's range of validity in time. In the quantum theory, the fluctuations $\delta A^a_i(x)$ cannot be neglected in general: While the expectation value $\langle \delta A^a_i(x)\rangle \equiv 0$ by definition, the correlation
$\langle \delta A^a_i(x) \delta A^b_j(y) \rangle$ cannot vanish identically because of the uncertainty relation. Starting with large macroscopic fields and all other modes in vacuum, where $Q^2/g^2 \sim \langle A^a_i(x) \rangle \langle A^b_j(x) \rangle \gg \langle \delta A^a_i(x) \delta A^b_j(y) \rangle \sim Q^2$ at initial times $x^0=y^0 = 0$, we have to investigate on which time scale fluctuations grow to become large enough such that they modify our result (\ref{eq:resultn5}). 

The initial growth of fluctuations is described by (\ref{eq:fluctuationequation}), evaluated for the macroscopic field configuration (\ref{eq:fieldconf}). Again carrying over the analysis of Ref.~\cite{Berges:2011sb} to our problem, we consider a Fourier expansion of the fluctuations and analyze which momentum modes dominate the growth of fluctuations. For the employed temporal gauge, the fluctuation equation can be written as a matrix equation of the form
\begin{align}
\partial_t^2(\delta A)=-\Omega^2[\bar{A}]\cdot(\delta A) \, .\label{eq:stability}
\end{align}
In fact, there are negative eigenvalues for $\Omega^2$ related to Nielsen--Olsen type instabilities~\cite{Nielsen:1978rm,Chang:1979tg}. 
In addition, there are parametric resonance instabilities arising from
the oscillatory behavior of the macroscopic field $\bar{A}(t)$, which are expected to be subleading
according to Ref.~\cite{Berges:2011sb}. Therefore, we proceed by computing the most negative eigenvalues of  $\Omega^2$ for constant $\bar{A} \sim Q$ to determine the characteristic exponential growth rates for fluctuations. 

From the three color times three spatial directions, $\Omega^2[\bar{A}]$ has a $9\times 9$ matrix structure. 
In the spatial momentum space, its nine eigenvalues depend on momentum $\bp$ only through its modulus $p=|\bp|$, 
and they read:
\begin{align}
\omega_{1/2}^2&=p^2\pm2|\mathcal{A}|\;p \, ,\label{eq:eigenvalues1}\\
\omega_{3/4/5/6}^2&=\mathcal{A}^2+\frac{1}{2} p^2\pm \frac{1}{2} \sqrt{(2\mathcal{A}^2+p^2)^2\pm8|\mathcal{A}|^3 p}\, ,\label{eq:eigenvalues2}
\end{align}
while $\omega_{7/8/9}^2$ are given by the roots of
\begin{align}
0=-4 \mathcal{A}^4 p^2 + (12 \mathcal{A}^4 + 4 \mathcal{A}^2 p^2 +
    p^4) \;x \nonumber\\- (8 \mathcal{A}^2 + 2 p^2)\; x^2 + x^3 \, ,
\end{align}
which are always non-negative. We find that (\ref{eq:eigenvalues1}) has negative eigenvalues for $0<p<2|\mathcal{A}|$, 
with the largest negative eigenvalue for $p_\star=|\mathcal{A}|$ given by $-\mathcal{A}^2$.
Similarly (\ref{eq:eigenvalues2}) yields negative eigenvalues, with the largest for $p_\star=(1+\sqrt{5})|\mathcal{A}|/2$ given by $-(\sqrt{5}-1)\mathcal{A}^2/2$. 
Since $\mathcal{A} \sim Q$, we conclude that the characteristic growth of fluctuations with momentum $p_\star \sim Q$ is described by an exponential behavior with rate $\gamma_\star \sim Q$. In spatial Fourier space, we therefore find for the fastest growing linear combination of fields
the parametric behavior
\begin{equation}
\langle \delta A\delta A \rangle (t,p_\star) \sim Q^{-1} \, e^{\gamma_\star t} \, .
\end{equation}

Next-to-leading order quantum corrections to the leading weak-coupling behavior of the fluctuation equation (\ref{eq:fluctuationequation}), both from gauge-field and fermion fluctuations, are proportional to $g^2$~(see e.g.~\cite{Kasper:2014uaa}). Parametrically, these quantum corrections are expected to become relevant once 
they have grown enough such that they can compensate for the small factor of $g^2$. Stated differently, they become relevant at the time $t_\Theta$ at which the dimensionless product  
\begin{equation}
g^2 Q \, \langle \delta A\delta A \rangle (t,p_\star )  \, \sim \, g^2 \, e^{\gamma_\star t} 
\end{equation}  
is of order unity, i.e.~at the time
\begin{equation}
t_\Theta \, \sim \, Q^{-1} \log (g^{-2}) \, .
\end{equation}
Before that time, the analytic estimate  (\ref{eq:resultn5}) for the anomalous charge generation dynamics may also be used to test real-time lattice simulation techniques that can be applied to more general out-of-equilibrium situations.

\subsection{Real-time lattice gauge theory simulations}

In this section, we go beyond the linear analysis by conducting classical-statistical lattice simulations for the pure gauge theory using standard procedures~\cite{Romatschke:2006nk,Berges:2007re,Kurkela:2012hp,Berges:2013eia,Gelis:2013rba}.
The system is defined by the lattice Hamiltonian for gauge fields
\begin{align}
H_g&=\frac{a^3}{2}\sum\limits_{\textbf{x},i}E^a_{i}(x)E^a_{i}(x)\nonumber\\&+\frac{2N_c}{g^2a}\sum\limits_{\textbf{x},i<j}\left( 1-\frac{1}{N_c}\text{Re\;Tr}\;U_{ij}(x)\right) , \label{eq:Hg}
\end{align}
where $a$ denotes the spacing of the isotropic spatial lattice, and $U_{ij} (x)$ is the spatial plaquette defined by
\begin{equation}
U_{ij} (x) = U_i (x) U_j (x+\hat{i}) U_i^\dagger (x+\hat{j}) U_j^\dagger (x). 
\end{equation}
Here $U_i (x) =\exp \left\{ igaA_i (x)\right\}$ is the link variable describing the gauge degrees of freedom on the lattice. 
While we discretize the space coordinates, time is treated as a continuum variable in this formulation.
We define the lattice magnetic field as
\begin{align}
B_i^a (x) = -\frac{\epsilon_{ijk} }{g a^2} \text{Im} \text{Tr} \left[ T^a U_{jk} (x)\right] \,.
\end{align}
This definition reduces to the continuum result $B_i=-\frac{1}{2}\epsilon_{ijk}F_{jk}$ as $a \rightarrow 0$. 

To simulate the instability beyond the linear analysis, 
the following initial fluctuations are added to the coherent field initial conditions (\ref{eq:AIC}):
\begin{gather}
\delta A_i^a (0,\bx ) = \sum_{\lambda =1,2} \frac{1}{V} \sum_{\bk} \frac{1}{\sqrt{2|\bk|}}
 \left[ \epsilon_{i, \bk}^{(\lambda)} c_{\lambda ,\bk}^a e^{i\bk \cdot \bx} +\text{c.c} \right] , \\
\delta E_i^a (0,\bx ) = -i \sum_{\lambda =1,2} \frac{1}{V} \sum_{\bk} \sqrt{\frac{|\bk |}{2}}
 \left[ \epsilon_{i, \bk}^{(\lambda)} c_{\lambda ,\bk}^a e^{i\bk \cdot \bx} -\text{c.c} \right] ,
\end{gather} 
where $\epsilon_{i, \bk}^{(\lambda)}$ is the transverse polarization vector\footnote{%
We restore the Gauss law in this construction
following \mbox{Ref.~\cite{Moore:1996qs}.}}. 
The ensemble average over random numbers $c_{\lambda ,\bk}^a$ is taken according to the variance
\begin{equation}
\langle c_{\lambda ,\bk}^a c_{\lambda^\prime ,\bk^\prime}^b \rangle
 = \delta_{\lambda,\lambda^\prime} \delta^{a,b} V \delta_{\bk ,\bk^\prime} .
\end{equation}  

\begin{figure}[t]
 \begin{center}
 \includegraphics[width=8.5cm]{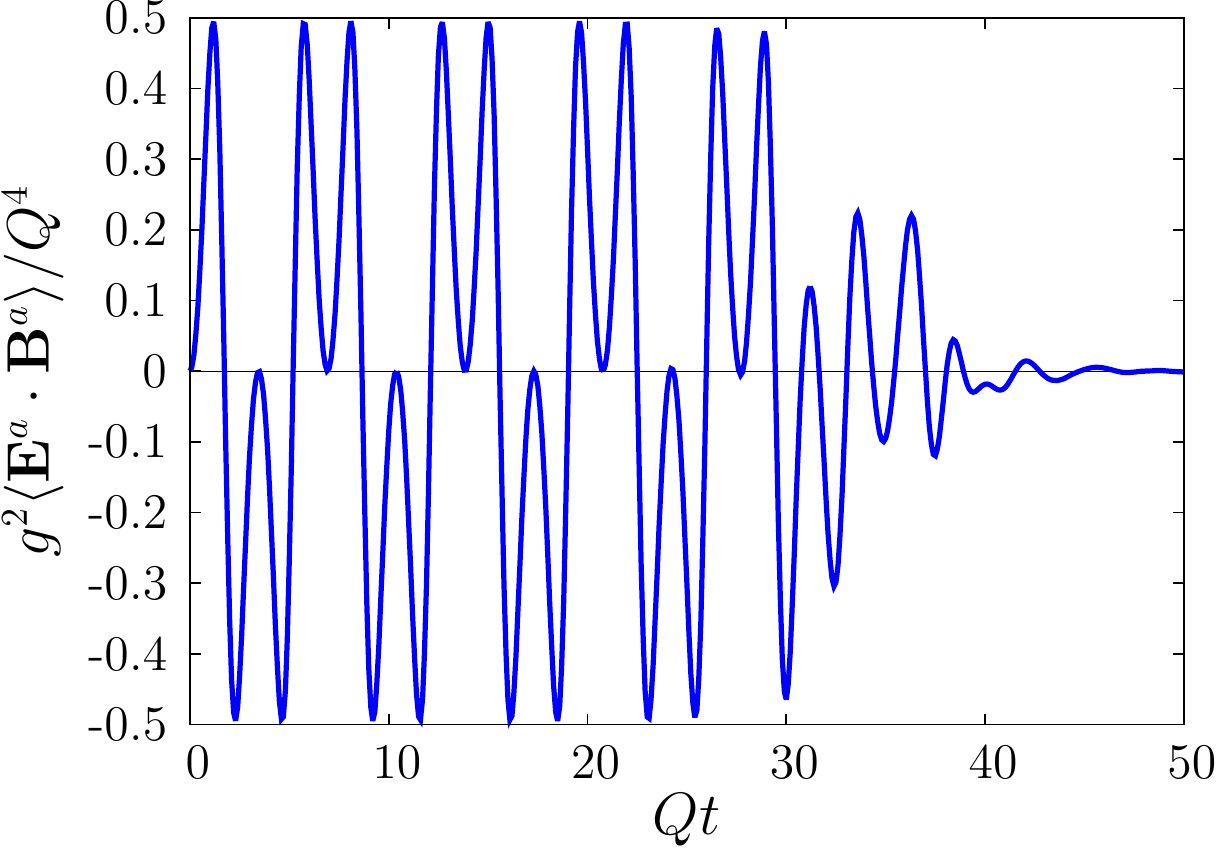} 
  \caption{Time dependence of $\langle \mathbf{E}^a \cd \mathbf{B}^a\rangle$ from classical-statistical lattice simulations. Comparison to Fig.~\ref{fig:EdotB} shows very good agreement with the linear analysis at early times, while the growing fluctuations become relevant around $t_\Theta$ leading to the significant changes observed.}
 \label{fig:EdotB_fluct} 
 \end{center}
\end{figure}

The non-perturbative lattice simulations take into account classical-statistical fluctuations up to arbitrary powers in $\delta A^a_i(t,\bx)$. As such, we expect agreement with the above analytic results for the linear approximation at early times, while deviations should occur around $t_\Theta$, when higher powers of $\delta A^a_i(t,\bx)$ become relevant. 
In Fig.~\ref{fig:EdotB_fluct}, we plot the time dependence of $\langle \mathbf{E}^a \cd \mathbf{B}^a\rangle$,
which is averaged over space-coordinates as well as random initial configurations. 
At early times, the effects of the small fluctuations are invisible and the result are in very good agreement with the analytical solution that is plotted in Fig.~\ref{fig:EdotB}. 
At later times around $t_\Theta$, however, the exponentially growing fluctuations cause decoherence 
of the uniform fields, and thus the ensemble average of  $\mathbf{E}^a \cd \mathbf{B}^a$ is diminished
significantly and approaches zero quickly afterwards as expected. The number of random initial configurations used in this computation is $N_\text{conf} =128$, where we have checked that sufficient convergence is obtained. The other parameters used for this computation are $g=10^{-3}$, $N_\text{latt} =64^3$ and $Qa =0.312$ for lattice spacing $a$.

When computing $\mathbf{E}^a \cdot \mathbf{B}^a$, one can use higher-order definitions of the electric field and the magnetic field with respect to lattice spacing;
e.g. the forward-backward averaged definition of the electric field and the clover-averaged definition of the magnetic field \cite{Moore:1996wn}.
We have numerically checked that for the configurations investigated in this work the naive and the higher order definitions of magnetic and electric fields agree.
We expect that the use of higher-order definitions is important for more inhomogeneous configurations. 

\begin{figure}[t]
 \begin{center}
 \includegraphics[width=8.5cm]{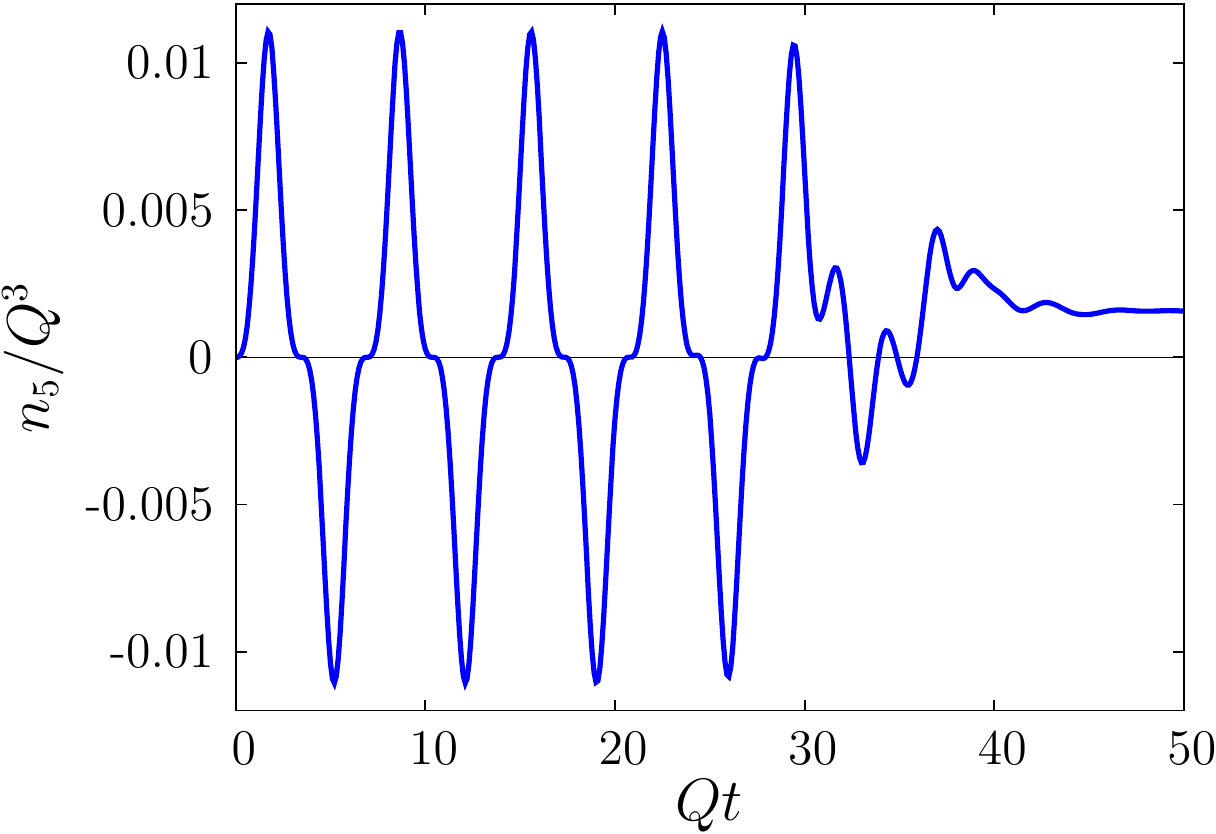} 
  \caption{Lattice simulation results for the evolution of the axial charge density in the massless limit. The early-time behavior agrees well with the analytic result (\ref{eq:resultn5}) drawn in Fig.~\ref{fig:analytic_n5}. A nonzero charge density is seen to persist even well beyond the decoherence time of the macroscopic gauge fields.
                   }
 \label{fig:n5_fluct} 
 \end{center}
\end{figure}

Figure \ref{fig:n5_fluct} shows the axial charge density for the massless case, which is obtained by integrating the space and ensemble average of $\mathbf{E}^a \cd \mathbf{B}^a$ over time. 
Again, the early time behavior agrees well with the analytical result~\eqref{eq:resultn5}. 
Although the macroscopic color-electric and magnetic fields approach zero quickly after a time around $t_\Theta$, a non-vanishing axial charge density is seen to persist for a much longer time after $t_\Theta$. This is possible because the axial charge density is determined by the integrated time history of $\langle \mathbf{E}^a \cd \mathbf{B}^a \rangle$. 
This observation indicates that coherent gauge fields very efficiently produce an axial charge density at early times, while part of the induced density persist to be present even well beyond characteristic decoherence times.

\section{Transient anomalous charge production: fermion sector} \label{sec:fermion}

\begin{figure}[t]
 \begin{center}
  \includegraphics[width=8.5cm]{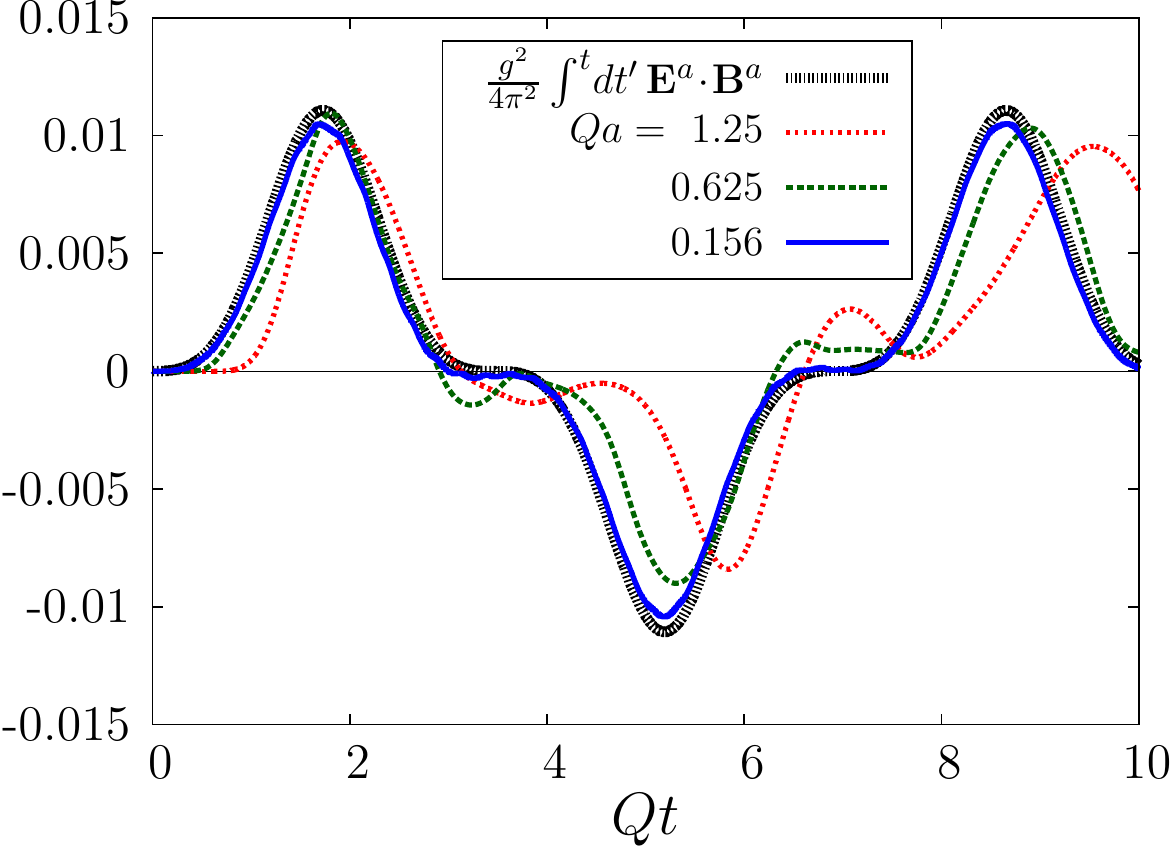} 
  \caption{Comparison of the Wilson term contribution $2r\int_0^t \! dt^\pr \text{Re} \langle \overline{\psi} i\gamma_5 W\psi \rangle /Q^3$ and the anomaly term $g^2 \int_0^t \! dt^\pr \langle \bE^a \cd \bB^a \rangle /(4\pi^2 Q^3)$ for three different values of the lattice spacing $a$ with fixed Wilson parameter $r=1$.}
 \label{fig:a-dep} 
 \end{center}
\end{figure}

\subsection{Axial anomaly with real-time Wilson fermions}
In this section, we investigate axial charge generation by using real-time lattice simulations with Wilson fermions~\cite{Aarts:1998td,Berges:2010zv,Saffin:2011kc,Saffin:2011kn,Hebenstreit:2013qxa,Kasper:2014uaa,Gelis:2015kya,Gelfand:2016prm,Buividovich:2015jfa}.  
This approach allows us to directly compute quark production and anomalous charge generation at leading order in the small coupling $g\ll 1$ for strong gauge fields \mbox{$A\sim Q/g$}. Consequently, we can use the lattice results to test the anomaly equation 
(\ref{eq:ABJ}) in this far-from-equilibrium situation by separately computing the fermion and gauge field terms on its left and right hand side. 

Starting from a homogeneous field configuration according to (\ref{eq:initcond}), for early times before $t_\Theta$ the gauge fields obey the classical Yang--Mills equations with vanishing color current, while the fermion field is determined through the Dirac equation in the background $SU(2)$ field:
\begin{equation} \label{eq:Dirac1}
\left(  i\gamma^0 \partial_0 +i\gamma^i D_i -m \right) \psi (x) = 0 \, , 
\end{equation}
in temporal gauge with $A_0^a =0$. Here we denote the spatial components of the covariant derivative by
\begin{equation}
D_i \psi = \left( \partial_i +igA_i^a T^a \right) \psi \, ,
\end{equation}
with the $SU(2)$ generators $T^a$. 

We can expand the field operator in terms of mode functions 
\begin{equation}
\psi (x) 
 = \sum_{s,c} \int \frac{d^3 p}{(2\pi)^3} \left[ \psi_{\bp ,s,c}^+ (x) a_{\bp ,s,c} 
   + \psi_{\bp ,s,c}^- (x) b_{\bp ,s,c}^\dagger \right] \, ,
\end{equation}
with $s$ being the spin and $c$ denoting the color label. Here 
$a_{\bp ,s,c} $ and $b_{\bp ,s,c} $ are annihilation operators for particles and antiparticles, respectively. Because the Dirac equation is linear, the mode functions obey the same Dirac equation as the field operator:
\begin{equation} \label{eq:Dirac2} 
\left(  i\gamma^0 \partial_0 +i\gamma^i D_i -m \right) \psi_{\bp ,s,c}^\pm (x) = 0 \, .  
\end{equation}
We consider for the initial state a perturbative vacuum, so that the initial condition for the mode functions at $t=0$ reads
\begin{gather} 
\psi_{\bp ,s,c}^+ (0 ,\bx ) = u(\bp ,s)\, \chi_c \, \frac{e^{-ip\cd x}}{\sqrt{2\omega_p}} \, ,\label{eq:initial_psi+} \\
\psi_{\bp ,s,c}^- (0 ,\bx ) = v(\bp ,s)\, \chi_c \, \frac{e^{+ip\cd x}}{\sqrt{2\omega_p}} \, , \label{eq:initial_psi-}
\end{gather}
with $\chi_c$ being a unit vector in color space. 
Once we obtain the mode functions by solving the equation \eqref{eq:Dirac2}, 
we can compute any observables expressed in terms of the field operator $\psi$. 

For the actual computations, we resort to a lattice discretization of the matter and gauge  fields. 
We add the following lattice Hamiltonian for the quark field to the Hamiltonian for the gauge field \eqref{eq:Hg}: 
\begin{align}
H_f&=a^3\sum\limits_{\textbf{x}}\Big\{m\bar{\psi}(x)\gamma^0\psi(x)\nonumber\\
&- \frac{1}{2a} \sum\limits_i \bar{\psi}(x)i\gamma^iU_{i}(x)\psi({x+\hat{i}})\nonumber\\
&+ \frac{1}{2a}\sum\limits_i \bar{\psi}(x)i\gamma^iU_i^\dagger(x-\hat{i}) \psi({x-\hat{i}})\Big\}.
\end{align}
The fermion doubling problem is regularized by adding a spatial Wilson term, which we will specify later.

\begin{figure}[t]
 \begin{center}
  \includegraphics[width=8.5cm]{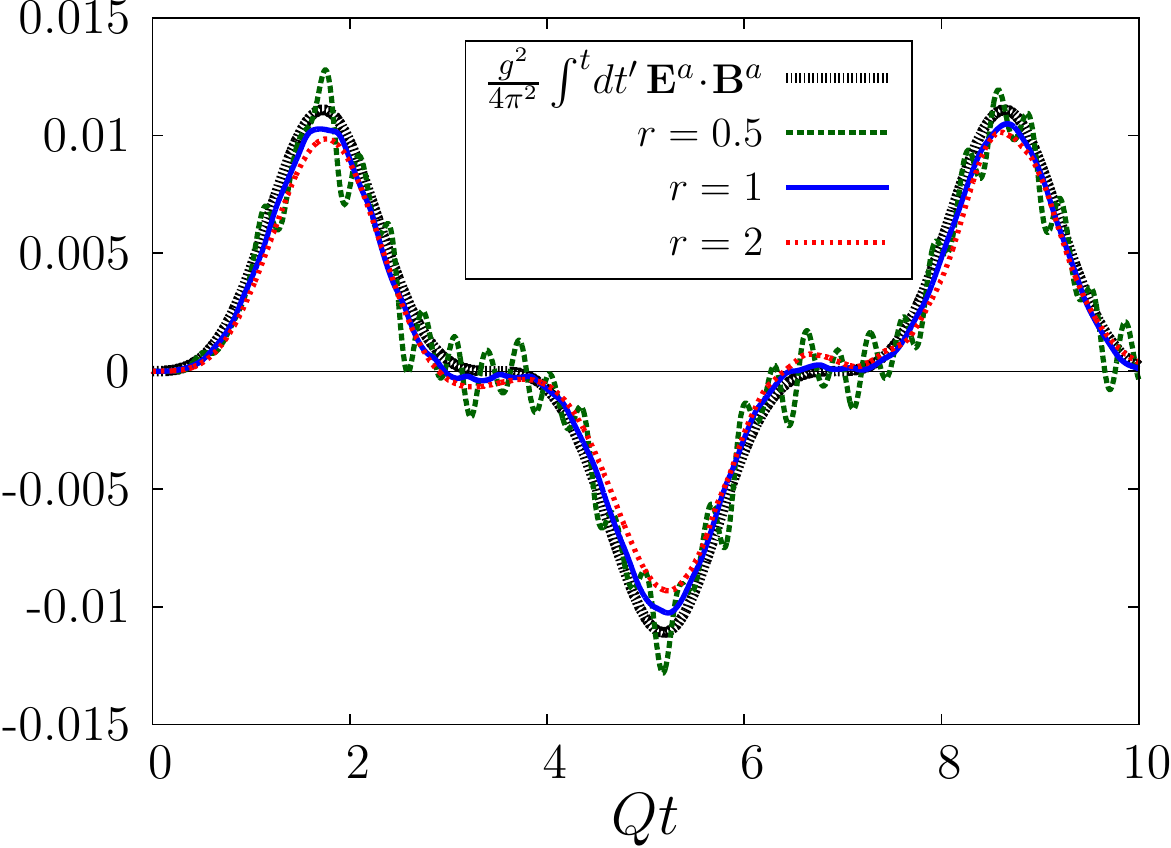} 
  \caption{Time evolution of the same quantities as in Fig.~\ref{fig:a-dep}, however, now the Wilson term contribution is shown for three different values of the Wilson parameter $r$ with fixed lattice spacing $Qa =0.208$.}
 \label{fig:r-dep} 
 \end{center}
\end{figure}

\begin{figure*}[t]
\begin{center}
\includegraphics[clip,width=8.5cm]{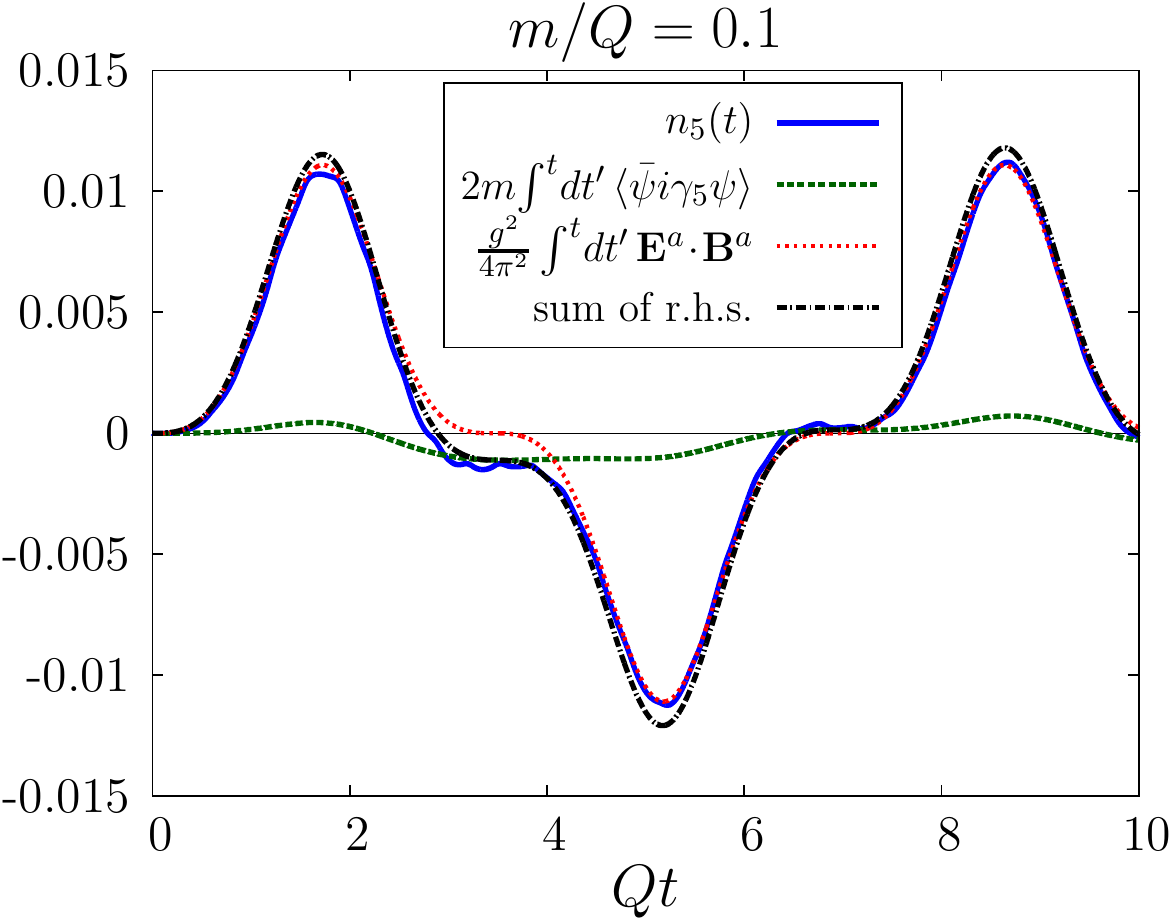} \ 
\includegraphics[clip,width=8.5cm]{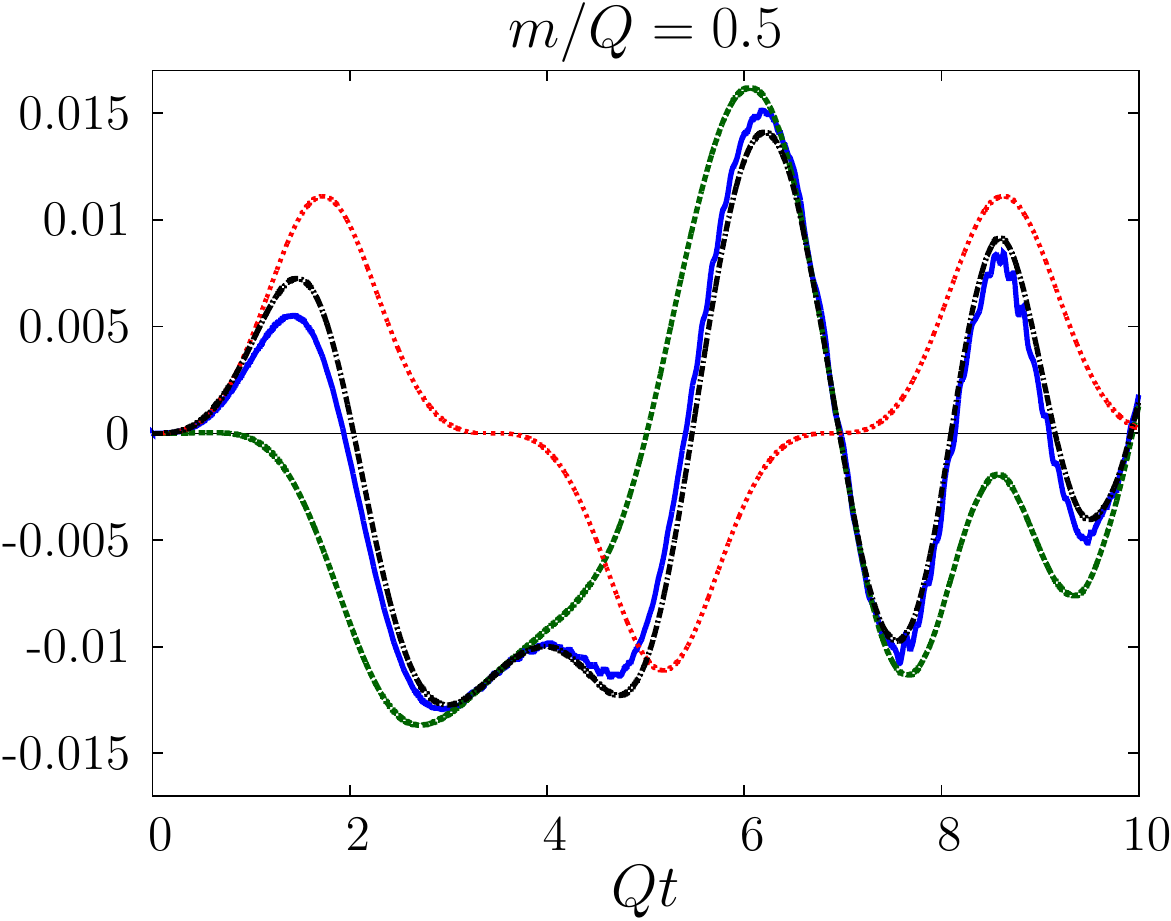} 
\end{center}
\caption{Time evolution of the various terms appearing in the anomaly equation \eqref{eq:intABJ} for two different fermion masses $m/Q=0.1$ (left) and for $m/Q=0.5$ (right). 
The sum of the two terms on the right hand side of \eqref{eq:intABJ} is plotted as well. Its agreement to $n_5$ within the expected accuracy for the employed lattice spacing provides a crucial validity check of the employed real-time regularization scheme.}
\label{fig:anom_eq}
\end{figure*}

The expectation value of both the axial charge density and the pseudo-scalar condensate are expressed in terms of the mode functions as
\begin{equation}
n_5 (t) = \frac{1}{V} \sum_{s,c} \sum_\bp \psi_{\bp ,s,c}^{-\, \dagger} (x) \gamma_5 \psi_{\bp ,s,c}^- (x) \, , 
\end{equation}
and
\begin{equation}
\langle \overline{\psi} (x) i\gamma_5 \psi (x) \rangle
 = \frac{1}{V} \sum_{s,c} \sum_\bp \psi_{\bp ,s,c}^{-\, \dagger} (x) i\gamma^0 \gamma_5 \psi_{\bp ,s,c}^- (x) \, , 
\end{equation}
respectively.

The realization of the axial anomaly on a lattice is non-trivial. In fact, it is closely related to the lattice fermion doubling problem and the anomaly is recovered by introducing a regulator term, removing the doubler fermions. As is well known, from the lattice Dirac equation \eqref{eq:Dirac1} one can easily compute the four-divergence of the axial current and obtain an anomaly-free equation. Correspondingly, if we numerically solve the Dirac equation \eqref{eq:Dirac2} with the naive lattice fermions, both $\langle \partial_\mu j_5^\mu \rangle$ and 
$2m \langle \overline{\psi} (x) i\gamma_5 \psi (x) \rangle$ are zero,
and thus the anomalous contributions cancel out.
In contrast, if one breaks the chiral symmetry explicitly by introducing a Wilson term, the axial anomaly is recovered by the continuum limit of this regulator, as has been studied in detail in Euclidean field theory~\cite{Karsten:1980wd,Nielsen:1980rz,Friedan:1982nk,Rothe:1998ba,Reisz:1999cma}.

In order to recover the anomaly in real-time simulations, typically a combination of a spatial Wilson term together with a suppression of possible temporal doublers using suitable initial conditions are employed~\cite{Saffin:2011kn,Saffin:2011kc}. 
The Dirac equation with the spatial Wilson term reads
\begin{equation} \label{DiracW}
\left( i\gamma^0 \partial_0 +i\gamma^i D_i -m \right) \psi (x) 
 +r W\psi (x) = 0 \, , 
\end{equation}
where $r$ is a positive constant and we have introduced an abbreviated notation
\begin{align}
W\psi (x) = \frac{1}{2a} \sum_{i=1}^3  
 \Big[ U_i (x)\psi (x+\hat{i} ) -2\psi (x)\nonumber\\ +U_i^\dagger (x-\hat{i} ) \psi (x-\hat{i} ) \Big] . 
\end{align}
For the specific case of a homogeneous background gauge field, one could in principle directly restrict the Brillouin zone to remove doublers as well, as long as this is done for the covariant (kinetic) momentum and hence in a gauge invariant way.\footnote{%
In this case, specific non-chiral observables, like the energy-momentum tensor and the charge current, can also be computed with a cutoff to the canonical momentum \cite{Tanji:2015ata}.} 
We comment on this possibility in the appendix.

Analyzing the anomaly on the lattice, it is helpful to point out that 
the relation \eqref{eq:intABJ} between the axial charge density, the time-integral of the pseudo-scalar condensate and of the anomaly term $\sim \langle \mathbf{E}^a \cd \mathbf{B}^a\rangle$ is not realized unless one takes the continuum limit. However, there exists 
a modified equation that is exactly satisfied on the lattice: 
\begin{equation} \label{eq:anomalyW2}
n_5 (t) = 
 2m\int_0^t \! dt^\pr \, \langle \overline{\psi} i\gamma_5 \psi \rangle  
 +2r\int_0^t \! dt^\pr \,\text{Re} \langle \overline{\psi} i\gamma_5 W\psi \rangle (t^\prime ) , 
\end{equation}
which can be derived from~\eqref{DiracW}. 
Comparing~\eqref{eq:intABJ} and \eqref{eq:anomalyW2}, one concludes that for $r \neq 0$ the Wilson term contribution is responsible for the anomaly term:
\begin{equation} \label{eq:WEB}
2r\, \text{Re} \langle \overline{\psi} i\gamma_5 W\psi \rangle 
\simeq 
\frac{g^2}{4\pi^2} \langle \mathbf{E}^a \cd \mathbf{B}^a \rangle\, , 
\end{equation}
which is expected to be accurate only in the continuum limit. 

Fig.~\ref{fig:a-dep} compares simulation results for the time-integrated left and right hand sides of (\ref{eq:WEB}) employing different values for the lattice spacing $a$. One observes that the relation (\ref{eq:WEB}) emerges for sufficiently small lattice spacing. The employed volume $V$ for these computations is $Q^3 V=20^3$, and we have employed $r=1$ for $m/Q=0.1$. In fact, in the continuum limit this result is insensitive to the precise value of $r \neq 0$ although $r$ apparently appears in (\ref{eq:WEB}) as an overall factor. In Fig.~\ref{fig:r-dep}, we show the same Wilson term contribution for different values of the Wilson parameter $r$ and fixed lattice spacing $Qa =0.208$, for $m/Q=0.1$ and $N_\text{latt} =96^3$. For the employed finite lattice spacing, small dependencies on $r$ and deviations from the anomaly term $\sim \int_0^t \! dt^\pr \langle \bE^a \cd \bB^a\rangle$ are still visible, which will also reflect the level of accuracy for the anomaly on the lattice in our calculations. 

After this preparatory analysis, we are now in a position to check the anomaly equation \eqref{eq:intABJ} by separately computing each of its terms. Since the results will depend on the explicit mixing of the different chiral components in the presence of a mass $m \neq 0$, we show in Fig.~\ref{fig:anom_eq} the evolution of the axial charge density, of the time-integral of the pseudo-scalar condensate and of the anomaly term for two different masses $m/Q=0.1$ and $m/Q=0.5$. The sum of $2m\int_0^t \! dt^\pr \, \langle \overline{\psi} i\gamma_5 \psi \rangle$ and of the anomaly term $g^2 \int_0^t \! dt^\pr \langle \bE^a \cd \bB^a \rangle/(4\pi^2)$ is also shown, since it has to agree to $n_5$ if the anomaly is accurately represented. The parameters used for these computations are
$N_\text{latt} =96^3$, $r=1$ and $Qa =0.208$ for $m/Q=0.1$, and $Qa =0.0625$ for $m/Q=0.5$.

For $m/Q=0.1$, one observes from Fig.~\ref{fig:anom_eq} that the anomaly term clearly dominates compared to the contributions from the pseudo-scalar condensate, while for $m/Q=0.5$ the pseudo-scalar term gives a larger contribution. 
In both cases, we find that the anomaly equation \eqref{eq:intABJ} is satisfied up to the expected accuracy for the employed lattice spacings. This provides an important consistency check for the employed real-time regularization with a spatial Wilson term. 

\begin{figure}[t]
 \begin{center}
  \includegraphics[width=8.7cm]{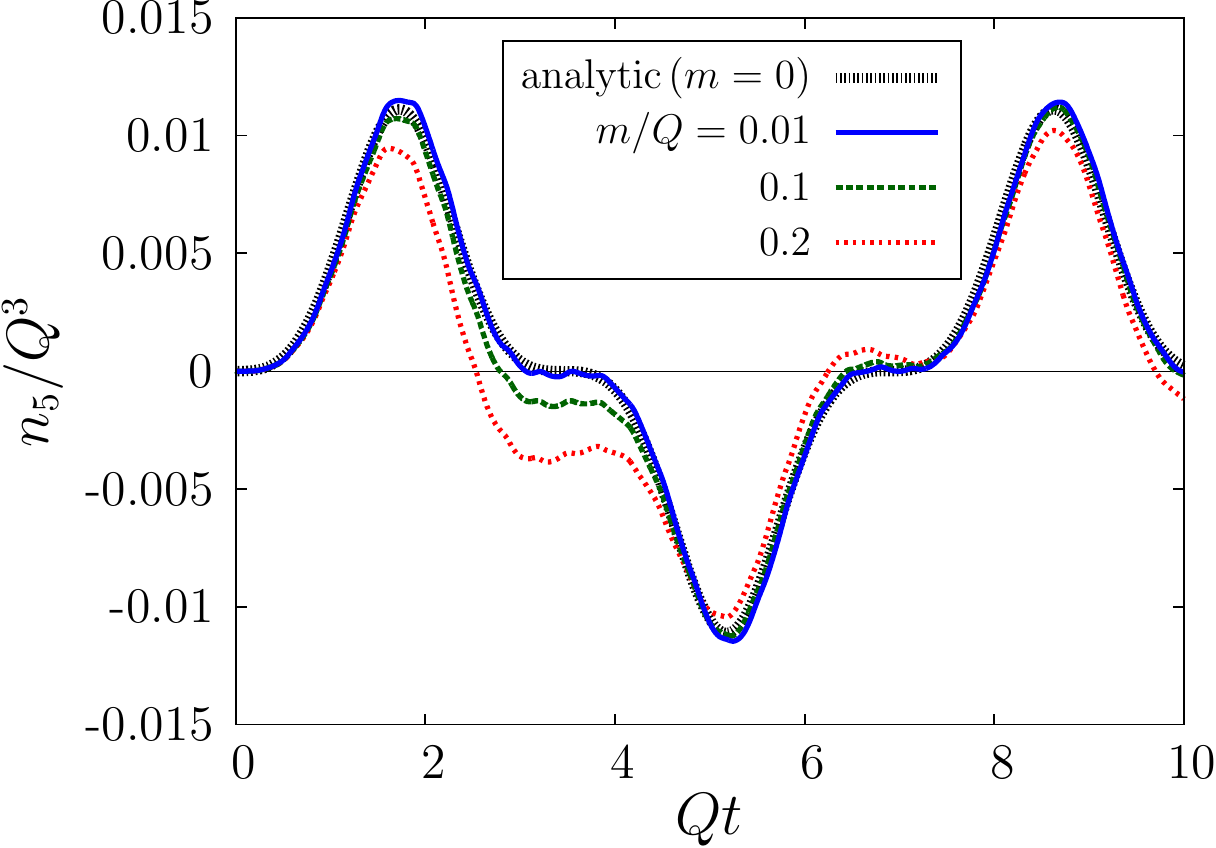} 
  \caption{Time dependence of the axial charge density. 
               The analytic result \eqref{eq:resultn5}, which is valid for the massless limit, and
               the results of the lattice simulations for different values of the quark mass are compared.
               The parameters used for the numerical computations are
               $N_\text{latt} =96^3$, $Qa =0.208$, and $r=1$. 
                }
 \label{fig:n5_mass} 
 \end{center}
\end{figure}

A further verification can be obtained from the comparison of the numerical lattice results for the axial charge density with the analytic solution \eqref{eq:resultn5}. Since the latter is only applicable to the massless case, we perform different lattice simulations with decreasing but nonzero fermion masses in order to be able to study numerically the approach to the massless limit.
Fig.~\ref{fig:n5_mass} displays the analytic $m=0$ curve along with lattice simulation results for three different values of the fermion mass: $m/Q=0.2$, $0.1$ and $0.01$. One observes that with lighter fermion masses the numerical results gets closer to the analytic curve \eqref{eq:resultn5}. In fact, Fig.~\ref{fig:n5_mass} exhibits a remarkably good agreement of the massless limit and the massive lattice results already for $m/Q=0.01$. 
This comparison represents a powerful demonstration that the axial anomaly is described by our real-time lattice simulations to very good accuracy.

\section{Conclusions} \label{sec:conclusion}

In this work we have investigated the out-of-equilibrium dynamics of anomalous quark production in two-color QCD. We have shown that the generation of a nonzero axial charge density can be described analytically for a class of initial conditions characterized by large coherent gauge fields motivated from the glasma picture. 
Employing real-time lattice simulations, we find that a transient anomalous charge density persists in this case even for times significantly exceeding the decoherence time of the macroscopic color-electric and magnetic fields.   
These findings can be very interesting for nonequilibrium phenomena such as the chiral magnetic effect during the very early stages of a relativistic heavy-ion collision, where finite-time effects may play a decisive role. 

Our combination of analytical and numerical results provides stringent tests of real-time representations of the axial anomaly in lattice QCD. We have investigated in detail the regularization of the fermion doubling problem using a spatial Wilson term. To this end, we considered first a modified anomaly equation that is exactly fulfilled on the lattice, and discussed the impact of a finite lattice spacing for computations of anomalous contributions in the continuum limit. 
This allowed us to check the anomaly equation by separately computing the different nonequilibrium fermion and gauge correlation functions entering that equation. In particular, we have confirmed the insensitivity of the real-time results to the specific choice of the Wilson parameter approaching the continuum limit.

The present work provides an important basis for more realistic simulations of anomalous nonequilibrium or transport processes in QCD related to heavy-ion collisions. Following along these lines, a wide range of dynamical phenomena can be addressed with ab initio calculations, from the intriguing interplay of non-Abelian and Abelian gauge fields underlying the chiral magnetic effect to possible chiral plasma instabilities~\cite{Akamatsu:2013pjd,Akamatsu:2015kau} followed into the far-from-equilibrium regime.  

\vspace{10pt}
The authors thank F.~Hebenstreit, V.~Kasper, M.~Mace, J.~Pawlowski, A.~Rothkopf, S.~Schlichting, and R.~Venugopalan  for  helpful discussions. 
N.M. thanks Brookhaven National Laboratory for hospitality while this work was completed. 
N.M.\ acknowledges support by the {\it Studienstiftung des Deutschen Volkes}. Part of this work was performed on the computational resource \mbox{ForHLR Phase I} funded by the Ministry of Science, Research and the Arts Baden-W\"urttemberg and the {\it Deutsche Forschungsgemeinschaft} (DFG).

\appendix
\section{Chiral anomaly and cutoff regularization} \label{sec:analytic_anomaly}

For the specific case of fermion dynamics in the presence of homogeneous background gauge fields considered in section~\ref{sec:fermion}, there is in principle an efficient alternative procedure to implement the anomaly on the lattice by restricting the Brillouin zone to remove doublers~\cite{Fukushima:2015tza}. 
In this appendix, by using analytic solutions of the Dirac equation for the example of quantum electrodynamics (QED), we will demonstrate that it is crucial for such a procedure to implement the corresponding momentum cutoff to preserve gauge invariance in order to describe the anomaly correctly. 

For QED in a uniform system with background electric field $\bE$ and magnetic field $\bB$, the anomaly equation reads
\begin{equation}
\partial_t n_5 (t) = \langle \overline{\psi} i\gamma_5 \psi \rangle
 +\frac{e^2}{2\pi^2} \bE \cd \bB\, , 
\end{equation}
where $e$ denotes the electromagnetic coupling.
In the following, we will verify this equation by using analytic solutions of the Dirac equation in the presence of a homogeneous background field that carries nonzero $\bE \cdot \bB$.
As an example of field configurations with nonzero $\bE \cdot \bB$ that are simple enough to access analytical solutions of the Dirac equation, 
we consider a Sauter-type pulsed electric field superposed by a constant magnetic field:
\begin{equation} \label{eq:Sauter+B}
\bE (t) = \frac{E}{\cosh ^2 (t/\tau)} {\mathbf e}_z , \hspace{10pt}
\bB = B {\mathbf e}_z , 
\end{equation}
where $E$, $B$, and the characteristic pulse time $\tau$ are constant, with ${\mathbf e}_z$ denoting a unit vector in the $z$-direction.
In the following, we assume $eE >0$ and $eB >0$. 

\subsection{Analytic solutions of the Dirac equation}
The electromagnetic fields \eqref{eq:Sauter+B} are described by the gauge potential
\begin{equation} \label{eq:gauge_Sauter}
A^\mu = \left( 0,-By,0,-E\tau \left[\tanh (t/\tau) +1\right] \right). 
\end{equation}
Under this gauge field, the mode functions are 
labeled by momenta $p_x$ and $p_z$, Landau level $n$ ($=0,1,2,\cdots $), 
and spin $s$ ($=\uparrow ,\downarrow$) as well as labels $+$ and $-$ that distinguish positive and negative energy solutions.
We employ the Dirac representation for the gamma matrices, and 
use the following basis to expand spinors:
\begin{equation}
\begin{split}
&\Gamma_1 = \frac{1}{\sqrt{2}} \begin{pmatrix} 1 \\ 0 \\ 1 \\ 0 \end{pmatrix} , \
  \Gamma_2 = \frac{1}{\sqrt{2}} \begin{pmatrix} 0 \\ 1 \\ 0 \\ -1 \end{pmatrix} , \\
&\Gamma_3 = \frac{1}{\sqrt{2}} \begin{pmatrix} 1 \\ 0 \\ -1 \\ 0 \end{pmatrix} , \
  \Gamma_4 = \frac{1}{\sqrt{2}} \begin{pmatrix} 0 \\ 1 \\ 0 \\ 1 \end{pmatrix} .
\end{split}
\end{equation}
The mode functions are
conveniently described by the following dimensionless parameters
\begin{gather}
\xi = \frac{1}{2} \left[ 1+\tanh (t/\tau )\right] , \\
\eta = \sqrt{\frac{2}{eB}} \left( eBy +p_x \right) , \\
\lambda = eE\tau ^2 , \\
\mu = \frac{\tau}{2} \sqrt{m^2 +eB(2n+1-s) +p_z^2} , \\
\nu = \frac{\tau}{2} \sqrt{m^2 +eB(2n+1-s) +(p_z +2eE\tau)^2} ,
\end{gather}
where $s=1$ for spin-up and $s=-1$ for spin-down. 
The mode functions are expressed in terms of these quantities as follows: 
\begin{widetext}
\begin{equation}
\begin{split} \label{eq:mf_S1}
\psi_{p_x ,p_z,n,\uparrow }^+ (x) 
 &= \left(\frac{eB}{\pi}\right)^{1/4} \sqrt{\frac{L}{n!}} 
      \frac{1}{\sqrt{4\mu (2\mu -p_z \tau)}} 
      \xi^{-i\mu} (1-\xi )^{-i\nu} \\
 &\hspace{10pt}  \times
      \left[ 2i \tilde\varphi^+ (\xi) D_n (\eta )\Gamma_3 
      -\sqrt{2eB} \tau \varphi^+ (\xi) nD_{n-1} (\eta ) \Gamma_2 +m\tau \varphi^+ (\xi) D_n (\eta ) \Gamma_1 \right] 
      e^{i(p_x x+p_z z)} , 
\end{split} 
\end{equation}
\begin{equation}
\begin{split} \label{eq:mf_S2}
\psi_{p_x ,p_z,n,\downarrow }^+ (x) 
 &= \left(\frac{eB}{\pi}\right)^{1/4} \sqrt{\frac{L}{n!}} 
      \frac{1}{\sqrt{4\mu (2\mu -p_z \tau)}} 
      \xi^{-i\mu} (1-\xi )^{-i\nu} \\
 &\hspace{10pt}  \times
      \left[ 2i \tilde\varphi^+ (\xi)  D_n (\eta )\Gamma_4 
      +\sqrt{2eB} \tau \varphi^+ (\xi) D_{n+1} (\eta ) \Gamma_1 +m\tau \varphi^+ (\xi) D_n (\eta ) \Gamma_2 \right] 
      e^{i(p_x x+p_z z)} , 
\end{split} 
\end{equation}
\begin{equation}
\begin{split} \label{eq:mf_S3}
\psi_{p_x ,p_z,n,\uparrow }^- (x) 
 &= \left(\frac{eB}{\pi}\right)^{1/4} \sqrt{\frac{L}{n!}} 
      \frac{1}{\sqrt{4\mu (2\mu +p_z \tau)}} 
      \xi^{+i\mu} (1-\xi )^{-i\nu} \\
 &\hspace{10pt}  \times
      \left[ -2i \tilde\varphi^- (\xi) D_n (\eta )\Gamma_3 
      +\sqrt{2eB} \tau \varphi^- (\xi) nD_{n-1} (\eta ) \Gamma_2 -m\tau \varphi^- (\xi) D_n (\eta ) \Gamma_1 \right] 
      e^{i(p_x x+p_z z)} , 
\end{split} 
\end{equation}
\begin{equation}
\begin{split} \label{eq:mf_S4} 
\psi_{p_x ,p_z,n,\downarrow }^- (x) 
 &= \left(\frac{eB}{\pi}\right)^{1/4} \sqrt{\frac{L}{n!}} 
      \frac{1}{\sqrt{4\mu (2\mu +p_z \tau)}} 
      \xi^{+i\mu} (1-\xi )^{-i\nu} \\
 &\hspace{10pt}  \times
      \left[ -2i \tilde\varphi^- (\xi) D_n (\eta )\Gamma_4 
      -\sqrt{2eB} \tau \varphi^- (\xi) D_{n+1} (\eta ) \Gamma_1 -m\tau \varphi^- (\xi) D_n (\eta ) \Gamma_2 \right] 
      e^{i(p_x x+p_z z)} , 
\end{split}
\end{equation}
where $L$ is the linear size of the system with volume $V=L^3$, and
$D_n (z)$ is the parabolic cylinder function. 
The functions $\varphi^\pm (\xi)$ and $\tilde\varphi^\pm (\xi)$ are represented by the hypergeometric function
${}_2F_1 (a,b;c;z)$ as follows:
\begin{align}
\varphi^\pm (\xi ) &= {}_2F_1 (\mp i\mu-i\nu-i\lambda  ,\mp i\mu-i\nu+i\lambda+1;1\mp 2i\mu;\xi) , \\
\begin{split}
\tilde\varphi^\pm (\xi ) 
 &= \xi^{\pm i\mu} (1-\xi )^{i\nu} \left[ \xi(1-\xi) \frac{d}{d\xi} +i\lambda \xi +\frac{i}{2} p_z \tau \right] \\
 &\hspace{10pt} \times
      \xi^{\mp i\mu} (1-\xi )^{-i\nu} {}_2F_1 (\mp i\mu-i\nu-i\lambda  ,\mp i\mu-i\nu+i\lambda+1;1\mp 2i\mu;\xi) \\
 &= (1\mp i\mu +i\nu+i\lambda ) {}_2F_1 (\mp i\mu-i\nu-i\lambda-1  ,\mp i\mu-i\nu+i\lambda+1;1\mp 2i\mu;\xi) \\
 &\hspace{10pt} 
      +\left[ \left( 1+2i\lambda \right) \xi -\left( 1+i\nu +i\lambda -\frac{i}{2} p_z \tau \right) \right] 
      {}_2F_1 (\mp i\mu-i\nu-i\lambda  ,\mp i\mu-i\nu+i\lambda+1;1\mp 2i\mu;\xi) . 
\end{split}
\end{align}
We note that the limits $t\to -\infty$ and $t\to +\infty$ correspond to $\xi \to 0$ and $\xi \to 1$, respectively. 
The mode functions $\psi_{p_x ,p_z,n,s}^+ (x) $ and $\psi_{p_x ,p_z,n,s}^- (x) $
satisfy the boundary condition such that at $t\to -\infty$
they approach the positive and negative energy solutions, respectively,
in a constant magnetic field. 
The mode functions are normalized by the inner product
\begin{equation}
\left( \psi_1 |\psi_2 \right)
 = \int \! d^3 x \, \psi_1^\dagger (t,\bx ) \psi_2 (t,\bx ) ,
\end{equation}
such that
\begin{gather}
\left( \psi_{p_x ,p_z,n,s}^+ \big| \psi_{p_x^\pr ,p_z^\pr ,n^\pr ,s^\pr }^+ \right) 
 = \delta_{s,s^\pr} L\delta_{n,n^\pr} (2\pi )^2 \delta (p_x- p_x^\pr ) \delta (p_z- p_z^\pr ) , \\
\left( \psi_{p_x ,p_z,n,s}^- \big| \psi_{p_x^\pr ,p_z^\pr ,n^\pr ,s^\pr }^- \right) 
 = \delta_{s,s^\pr} L\delta_{n,n^\pr} (2\pi )^2 \delta (p_x- p_x^\pr ) \delta (p_z- p_z^\pr ) , \\
\left( \psi_{p_x ,p_z,n,s}^+ \big| \psi_{p_x^\pr ,p_z^\pr ,n^\pr ,s^\pr }^- \right) 
 = \left( \psi_{p_x ,p_z,n,s}^- \big| \psi_{p_x^\pr ,p_z^\pr ,n^\pr ,s^\pr }^+ \right) = 0 . 
\end{gather}

In terms of the mode functions, the fermion field operator $\psi$ is expanded as
\begin{equation}
\psi (x) 
  = \sum_s \frac{1}{L} \sum_{n=0}^\infty \int \! \frac{dp_x}{2\pi} \int \! \frac{dp_z}{2\pi} 
     \left[ \psi_{p_x ,p_z,n,s}^+ (x) a_{p_x ,p_z,n,s} +\psi_{p_x ,p_z,n,s}^- (x) b_{p_x ,p_z,n,s}^\dagger \right] .
\end{equation}
The creation and annihilation operators satisfy
\begin{equation}
\begin{split}
\left\{ a_{p_x ,p_z,n,s} ,a_{p_x^\pr ,p_z^\pr,n^\pr,s^\pr}^\dagger \right\}
  = \left\{ b_{p_x ,p_z,n,s} ,b_{p_x^\pr ,p_z^\pr,n^\pr,s^\pr}^\dagger \right\} 
  = \delta_{s,s^\pr} L\delta_{n,n^\pr} (2\pi)^2 \delta (p_x-p_x^\pr ) \delta (p_z-p_z^\pr ) . 
\end{split}
\end{equation}

\subsection{Verification of the anomaly equation}
The vacuum expectation of the chiral charge density, $n_5 (t)$, is expressed by the mode functions as
\begin{align}
n_5 (t) 
 &= \sum_{s} \frac{1}{L} \sum_{n=0}^\infty \int \! \frac{dp_x}{2\pi} \int \! \frac{dp_z}{2\pi} 
      \psi_{p_x ,p_z,n,s}^{-\, \dagger} (x) \gamma_5 \psi_{p_x ,p_z,n,s}^- (x) .
\end{align}
\end{widetext}
After substituting the explicit forms of the mode functions (\ref{eq:mf_S3}-\ref{eq:mf_S4}), 
we can first execute the $p_x$-integral by using 
\begin{equation}
\int_{-\infty}^{\infty} \! dx \left[ D_n (x) \right]^2 = \sqrt{2\pi} n! .
\end{equation}
We note that the $p_x$-integration is finite without a cutoff. 
After the integration, 
it turns out that the contribution of the mode with $(n+1,\uparrow)$
and that with $(n,\downarrow)$ cancel each other. 
As a consequence, only the lowest mode $(n=0,s=\uparrow)$ contributes to the chiral charge, and one obtains:
\begin{align}
n_5 (t) 
 &= \frac{eB}{4\pi^2} \int \! dp_z \frac{1}{2\sqrt{m^2 +p_z^2} (\sqrt{m^2 +p_z^2} +p_z)} \notag \\
 &\hspace{10pt} \times
      \left[ -\frac{4}{\tau^2} \left| \tilde\varphi^- (\xi)\right|^2 +m^2 \left| \varphi^- (\xi)\right|^2 \right]_{n=0,s=\uparrow} . 
\label{eq:analytic_n5}
\end{align}
In a similar way, we can compute the pseudo-scalar condensate:
\begin{align}
\langle \overline{\psi} i\gamma_5 \psi \rangle
 &= \frac{eB}{4\pi^2} \int \! dp_z 
      \frac{1}{2\sqrt{m^2 +p_z^2} (\sqrt{m^2 +p_z^2}+p_z)} \notag \\
 &\hspace{10pt} \times
      \frac{4m}{\tau}  \text{Re} \left[ \varphi^{-\, *} (\xi) \tilde\varphi^- (\xi ) \right]_{n=0,s=\uparrow} . 
\end{align}
The right hand side of \eqref{eq:analytic_n5} depends on time only through $\xi $.
After some algebra, one finds that
\begin{equation}
\frac{\partial}{\partial t} \left| \varphi^- (\xi)\right|^2 
 = \frac{4}{\tau } \text{Re} \left[ \varphi^{-\, *} (\xi) \tilde\varphi^- (\xi ) \right] , 
\end{equation}
and
\begin{equation}
\frac{\partial}{\partial t} \left| \tilde\varphi^- (\xi)\right|^2 
 = -\tau m^2  \text{Re} \left[ \varphi^{-\, *} (\xi) \tilde\varphi^- (\xi ) \right] .
\end{equation}
Collecting all these results, we finally arrive at
\begin{equation}
\begin{split}
\partial_t n_5 
 &= 2m \langle \overline{\psi} i\gamma_5 \psi \rangle \, , 
 \label{eq:n5wa}
\end{split}
\end{equation}
which so far does not contain the anomaly term. 

In the diagrammatic derivation of the axial anomaly, it is crucial to regularize a divergent integral in a gauge-invariant way.
Also in our calculation, we need to regularize the integral in \eqref{eq:analytic_n5} to obtain the anomaly term.
In fact, the integrand of \eqref{eq:analytic_n5} does not fall off at $p_z \to \pm \infty$. 
By using the asymptotic expansion of the hypergeometric function
${}_2F_1 (a,b;c;z)$ for large $|c|$ \cite{abramowitz1965handbook}, 
one finds that 
\begin{equation} \label{eq:integrand}
\left\{ \text{integrand of \eqref{eq:analytic_n5}} \right\}
 \approx 
  \begin{cases}
   -1 & (p_z \to +\infty ) \\ 
   +1 & (p_z \to -\infty ) . 
  \end{cases}
\end{equation}
To regularize this divergent integral, one may 
naively introduce a cutoff for $p_z$ as
\begin{equation}
\int_{-\infty}^{+\infty} \! dp_z \longrightarrow \int_{-\Lambda}^{+\Lambda} \! dp_z \, 
\end{equation}
to see that this does not alter the result (\ref{eq:n5wa}). 
The reason why the anomaly term is not obtained is that 
it introduces the cutoff for the canonical momentum. 
The canonical momentum $\bp_\text{can}$ is related to the kinetic momentum $\bp_\text{kin}$ as
\begin{equation} \label{cano-kine}
\bp_\text{can} = \bp_\text{kin} +e{\mathbf A} \, . 
\end{equation}
While here the kinetic momentum is a gauge-invariant quantity, 
the canonical momentum is gauge-dependent.
In a translational-invariant system, the canonical momentum is a constant of motion,
and thus it is associated with a plane wave factor $e^{i\bp \cd \bx}$. 
Therefore, the momentum $p_z$ appearing in \eqref{eq:analytic_n5} is a canonical momentum. 
Since the canonical momentum is not a gauge-invariant quantity, putting a cutoff breaks gauge invariance.
In order to regularize the integral keeping the gauge invariance, 
we need to introduce a cutoff for the kinetic momentum.
Because of the relation \eqref{cano-kine}, putting a cutoff $\pm \Lambda$ to the kinetic momentum 
amounts to putting a time-dependent cutoff $\pm \Lambda +eA^3 (t)$ to the canonical momentum:
\begin{equation}
\int_{-\infty}^{+\infty} \! dp_z \longrightarrow \int_{-\Lambda +eA^3(t)}^{+\Lambda +eA^3(t)} \! dp_z \, .\label{eq:integral_reg}
\end{equation}
Thanks to this time-dependent cutoff, $\partial_t n_5  (t)$ acquires the anomaly term:
\begin{align}
\partial_t n_5   
 &= \frac{eB}{4\pi^2} 
      \partial_t \int_{-\Lambda +eA^3(t)}^{+\Lambda +eA^3(t)} \! dp_z 
      \biggl\{ \cdots \biggr\} \notag \\
 &= 2m\langle \overline{\psi} i\gamma_5 \psi \rangle
      +\frac{e^2 B}{4\pi^2} \frac{dA^3}{dt} 
      \biggl[ \cdots \biggr]_{p_z = -\Lambda +eA^3(t)}^{p_z = \Lambda +eA^3(t)} \notag \\
 &= 2m\langle \overline{\psi} i\gamma_5 \psi \rangle
      +\frac{e^2 }{2\pi^2} \bE  \cd \bB \, ,
\end{align}
where we have used $E_z = -dA^3/dt$ 
and \eqref{eq:integrand}.

In the massless case the gauge invariant regularization of \eqref{eq:integral_reg} can also be seen in the context of spectral flow (see e.g. \cite{Klinkhamer:2003hz}). The use of the covariant momentum corresponds to a time dependent rearrangement of the eigenvalues of the Hamiltonian. In the case of nonzero $\mathbf{E}\cdot\mathbf{B}$, the dispersion relation of the fermions is altered in such a way that the rearrangement is different for left and right handed particles and thus a net chiral charge is generated. 

We have demonstrated how the anomaly term appears from the gauge-invariant cutoff regularization. 
However, such a computation applies only to the specific case of an Abelian and uniform background gauge field. 
For non-Abelian and/or inhomogeneous gauge fields, the relation between kinetic and canonical momentum
becomes ambiguous. 
In that case, the lattice regularization with the Wilson term method provides a powerful way to describe the axial anomaly, 
as discussed in the main text.
Here we note that the derivative appearing in the Wilson term is the covariant derivative, 
and the covariant derivative corresponds to the kinetic momentum,
${\boldsymbol D} e^{i\bp_\text{can} \cd \bx} =i\bp_\text{kin} e^{i\bp_\text{can} \cd \bx}$.


\begin{thebibliography}{52}%
\makeatletter
\providecommand \@ifxundefined [1]{%
 \@ifx{#1\undefined}
}%
\providecommand \@ifnum [1]{%
 \ifnum #1\expandafter \@firstoftwo
 \else \expandafter \@secondoftwo
 \fi
}%
\providecommand \@ifx [1]{%
 \ifx #1\expandafter \@firstoftwo
 \else \expandafter \@secondoftwo
 \fi
}%
\providecommand \natexlab [1]{#1}%
\providecommand \enquote  [1]{``#1''}%
\providecommand \bibnamefont  [1]{#1}%
\providecommand \bibfnamefont [1]{#1}%
\providecommand \citenamefont [1]{#1}%
\providecommand \href@noop [0]{\@secondoftwo}%
\providecommand \href [0]{\begingroup \@sanitize@url \@href}%
\providecommand \@href[1]{\@@startlink{#1}\@@href}%
\providecommand \@@href[1]{\endgroup#1\@@endlink}%
\providecommand \@sanitize@url [0]{\catcode `\\12\catcode `\$12\catcode
  `\&12\catcode `\#12\catcode `\^12\catcode `\_12\catcode `\%12\relax}%
\providecommand \@@startlink[1]{}%
\providecommand \@@endlink[0]{}%
\providecommand \url  [0]{\begingroup\@sanitize@url \@url }%
\providecommand \@url [1]{\endgroup\@href {#1}{\urlprefix }}%
\providecommand \urlprefix  [0]{URL }%
\providecommand \Eprint [0]{\href }%
\providecommand \doibase [0]{http://dx.doi.org/}%
\providecommand \selectlanguage [0]{\@gobble}%
\providecommand \bibinfo  [0]{\@secondoftwo}%
\providecommand \bibfield  [0]{\@secondoftwo}%
\providecommand \translation [1]{[#1]}%
\providecommand \BibitemOpen [0]{}%
\providecommand \bibitemStop [0]{}%
\providecommand \bibitemNoStop [0]{.\EOS\space}%
\providecommand \EOS [0]{\spacefactor3000\relax}%
\providecommand \BibitemShut  [1]{\csname bibitem#1\endcsname}%
\let\auto@bib@innerbib\@empty
\bibitem [{\citenamefont {Kuzmin}\ \emph {et~al.}(1985)\citenamefont {Kuzmin},
  \citenamefont {Rubakov},\ and\ \citenamefont {Shaposhnikov}}]{Kuzmin:1985mm}%
  \BibitemOpen
  \bibfield  {author} {\bibinfo {author} {\bibfnamefont {V.~A.}\ \bibnamefont
  {Kuzmin}}, \bibinfo {author} {\bibfnamefont {V.~A.}\ \bibnamefont {Rubakov}},
  \ and\ \bibinfo {author} {\bibfnamefont {M.~E.}\ \bibnamefont
  {Shaposhnikov}},\ }\href {\doibase 10.1016/0370-2693(85)91028-7} {\bibfield
  {journal} {\bibinfo  {journal} {Phys. Lett.}\ }\textbf {\bibinfo {volume}
  {B155}},\ \bibinfo {pages} {36} (\bibinfo {year} {1985})}\BibitemShut
  {NoStop}%
\bibitem [{\citenamefont {Arnold}\ and\ \citenamefont
  {McLerran}(1987)}]{Arnold:1987mh}%
  \BibitemOpen
  \bibfield  {author} {\bibinfo {author} {\bibfnamefont {P.~B.}\ \bibnamefont
  {Arnold}}\ and\ \bibinfo {author} {\bibfnamefont {L.~D.}\ \bibnamefont
  {McLerran}},\ }\href {\doibase 10.1103/PhysRevD.36.581} {\bibfield  {journal}
  {\bibinfo  {journal} {Phys. Rev.}\ }\textbf {\bibinfo {volume} {D36}},\
  \bibinfo {pages} {581} (\bibinfo {year} {1987})}\BibitemShut {NoStop}%
\bibitem [{\citenamefont {Dine}\ and\ \citenamefont
  {Kusenko}(2003)}]{Dine:2003ax}%
  \BibitemOpen
  \bibfield  {author} {\bibinfo {author} {\bibfnamefont {M.}~\bibnamefont
  {Dine}}\ and\ \bibinfo {author} {\bibfnamefont {A.}~\bibnamefont {Kusenko}},\
  }\href {\doibase 10.1103/RevModPhys.76.1} {\bibfield  {journal} {\bibinfo
  {journal} {Rev. Mod. Phys.}\ }\textbf {\bibinfo {volume} {76}},\ \bibinfo
  {pages} {1} (\bibinfo {year} {2003})}\BibitemShut {NoStop}%
\bibitem [{\citenamefont {Kharzeev}\ \emph {et~al.}(2008)\citenamefont
  {Kharzeev}, \citenamefont {McLerran},\ and\ \citenamefont
  {Warringa}}]{Kharzeev:2007jp}%
  \BibitemOpen
  \bibfield  {author} {\bibinfo {author} {\bibfnamefont {D.~E.}\ \bibnamefont
  {Kharzeev}}, \bibinfo {author} {\bibfnamefont {L.~D.}\ \bibnamefont
  {McLerran}}, \ and\ \bibinfo {author} {\bibfnamefont {H.~J.}\ \bibnamefont
  {Warringa}},\ }\href {\doibase 10.1016/j.nuclphysa.2008.02.298} {\bibfield
  {journal} {\bibinfo  {journal} {Nucl. Phys.}\ }\textbf {\bibinfo {volume}
  {A803}},\ \bibinfo {pages} {227} (\bibinfo {year} {2008})}\BibitemShut
  {NoStop}%
\bibitem [{\citenamefont {Fukushima}\ \emph {et~al.}(2008)\citenamefont
  {Fukushima}, \citenamefont {Kharzeev},\ and\ \citenamefont
  {Warringa}}]{Fukushima:2008xe}%
  \BibitemOpen
  \bibfield  {author} {\bibinfo {author} {\bibfnamefont {K.}~\bibnamefont
  {Fukushima}}, \bibinfo {author} {\bibfnamefont {D.~E.}\ \bibnamefont
  {Kharzeev}}, \ and\ \bibinfo {author} {\bibfnamefont {H.~J.}\ \bibnamefont
  {Warringa}},\ }\href {\doibase 10.1103/PhysRevD.78.074033} {\bibfield
  {journal} {\bibinfo  {journal} {Phys. Rev.}\ }\textbf {\bibinfo {volume}
  {D78}},\ \bibinfo {pages} {074033} (\bibinfo {year} {2008})}\BibitemShut
  {NoStop}%
\bibitem [{\citenamefont {Fukushima}\ \emph {et~al.}(2010)\citenamefont
  {Fukushima}, \citenamefont {Kharzeev},\ and\ \citenamefont
  {Warringa}}]{Fukushima:2010vw}%
  \BibitemOpen
  \bibfield  {author} {\bibinfo {author} {\bibfnamefont {K.}~\bibnamefont
  {Fukushima}}, \bibinfo {author} {\bibfnamefont {D.~E.}\ \bibnamefont
  {Kharzeev}}, \ and\ \bibinfo {author} {\bibfnamefont {H.~J.}\ \bibnamefont
  {Warringa}},\ }\href {\doibase 10.1103/PhysRevLett.104.212001} {\bibfield
  {journal} {\bibinfo  {journal} {Phys. Rev. Lett.}\ }\textbf {\bibinfo
  {volume} {104}},\ \bibinfo {pages} {212001} (\bibinfo {year}
  {2010})}\BibitemShut {NoStop}%
\bibitem [{\citenamefont {Kharzeev}\ \emph {et~al.}(2016)\citenamefont
  {Kharzeev}, \citenamefont {Liao}, \citenamefont {Voloshin},\ and\
  \citenamefont {Wang}}]{Kharzeev:2015znc}%
  \BibitemOpen
  \bibfield  {author} {\bibinfo {author} {\bibfnamefont {D.~E.}\ \bibnamefont
  {Kharzeev}}, \bibinfo {author} {\bibfnamefont {J.}~\bibnamefont {Liao}},
  \bibinfo {author} {\bibfnamefont {S.~A.}\ \bibnamefont {Voloshin}}, \ and\
  \bibinfo {author} {\bibfnamefont {G.}~\bibnamefont {Wang}},\ }\href {\doibase
  10.1016/j.ppnp.2016.01.001} {\bibfield  {journal} {\bibinfo  {journal} {Prog.
  Part. Nucl. Phys.}\ }\textbf {\bibinfo {volume} {88}},\ \bibinfo {pages} {1}
  (\bibinfo {year} {2016})}\BibitemShut {NoStop}%
\bibitem [{\citenamefont {Sakharov}(1967)}]{Sakharov:1967dj}%
  \BibitemOpen
  \bibfield  {author} {\bibinfo {author} {\bibfnamefont {A.~D.}\ \bibnamefont
  {Sakharov}},\ }\href {\doibase 10.1070/PU1991v034n05ABEH002497} {\bibfield
  {journal} {\bibinfo  {journal} {Pisma Zh. Eksp. Teor. Fiz.}\ }\textbf
  {\bibinfo {volume} {5}},\ \bibinfo {pages} {32} (\bibinfo {year} {1967})},\
  \bibinfo {note} {[Usp. Fiz. Nauk \textbf{161}, 61 (1991)]}\BibitemShut
  {NoStop}%
\bibitem [{\citenamefont {Lappi}\ and\ \citenamefont
  {McLerran}(2006)}]{Lappi:2006fp}%
  \BibitemOpen
  \bibfield  {author} {\bibinfo {author} {\bibfnamefont {T.}~\bibnamefont
  {Lappi}}\ and\ \bibinfo {author} {\bibfnamefont {L.}~\bibnamefont
  {McLerran}},\ }\href {\doibase 10.1016/j.nuclphysa.2006.04.001} {\bibfield
  {journal} {\bibinfo  {journal} {Nucl. Phys.}\ }\textbf {\bibinfo {volume}
  {A772}},\ \bibinfo {pages} {200} (\bibinfo {year} {2006})}\BibitemShut
  {NoStop}%
\bibitem [{\citenamefont {Gelis}\ \emph {et~al.}(2010)\citenamefont {Gelis},
  \citenamefont {Iancu}, \citenamefont {Jalilian-Marian},\ and\ \citenamefont
  {Venugopalan}}]{Gelis:2010nm}%
  \BibitemOpen
  \bibfield  {author} {\bibinfo {author} {\bibfnamefont {F.}~\bibnamefont
  {Gelis}}, \bibinfo {author} {\bibfnamefont {E.}~\bibnamefont {Iancu}},
  \bibinfo {author} {\bibfnamefont {J.}~\bibnamefont {Jalilian-Marian}}, \ and\
  \bibinfo {author} {\bibfnamefont {R.}~\bibnamefont {Venugopalan}},\ }\href
  {\doibase 10.1146/annurev.nucl.010909.083629} {\bibfield  {journal} {\bibinfo
   {journal} {Ann. Rev. Nucl. Part. Sci.}\ }\textbf {\bibinfo {volume} {60}},\
  \bibinfo {pages} {463} (\bibinfo {year} {2010})}\BibitemShut {NoStop}%
\bibitem [{\citenamefont {Arnold}\ \emph {et~al.}(1997)\citenamefont {Arnold},
  \citenamefont {Son},\ and\ \citenamefont {Yaffe}}]{Arnold:1996dy}%
  \BibitemOpen
  \bibfield  {author} {\bibinfo {author} {\bibfnamefont {P.~B.}\ \bibnamefont
  {Arnold}}, \bibinfo {author} {\bibfnamefont {D.}~\bibnamefont {Son}}, \ and\
  \bibinfo {author} {\bibfnamefont {L.~G.}\ \bibnamefont {Yaffe}},\ }\href
  {\doibase 10.1103/PhysRevD.55.6264} {\bibfield  {journal} {\bibinfo
  {journal} {Phys. Rev.}\ }\textbf {\bibinfo {volume} {D55}},\ \bibinfo {pages}
  {6264} (\bibinfo {year} {1997})}\BibitemShut {NoStop}%
\bibitem [{\citenamefont {Moore}\ and\ \citenamefont
  {Turok}(1997)}]{Moore:1997cr}%
  \BibitemOpen
  \bibfield  {author} {\bibinfo {author} {\bibfnamefont {G.~D.}\ \bibnamefont
  {Moore}}\ and\ \bibinfo {author} {\bibfnamefont {N.}~\bibnamefont {Turok}},\
  }\href {\doibase 10.1103/PhysRevD.56.6533} {\bibfield  {journal} {\bibinfo
  {journal} {Phys. Rev.}\ }\textbf {\bibinfo {volume} {D56}},\ \bibinfo {pages}
  {6533} (\bibinfo {year} {1997})}\BibitemShut {NoStop}%
\bibitem [{\citenamefont {Saffin}\ and\ \citenamefont
  {Tranberg}(2012)}]{Saffin:2011kn}%
  \BibitemOpen
  \bibfield  {author} {\bibinfo {author} {\bibfnamefont {P.~M.}\ \bibnamefont
  {Saffin}}\ and\ \bibinfo {author} {\bibfnamefont {A.}~\bibnamefont
  {Tranberg}},\ }\href {\doibase 10.1007/JHEP02(2012)102} {\bibfield  {journal}
  {\bibinfo  {journal} {JHEP}\ }\textbf {\bibinfo {volume} {02}},\ \bibinfo
  {pages} {102} (\bibinfo {year} {2012})}\BibitemShut {NoStop}%
\bibitem [{\citenamefont {Mace}\ \emph {et~al.}(2016)\citenamefont {Mace},
  \citenamefont {Schlichting},\ and\ \citenamefont
  {Venugopalan}}]{Mace:2016svc}%
  \BibitemOpen
  \bibfield  {author} {\bibinfo {author} {\bibfnamefont {M.}~\bibnamefont
  {Mace}}, \bibinfo {author} {\bibfnamefont {S.}~\bibnamefont {Schlichting}}, \
  and\ \bibinfo {author} {\bibfnamefont {R.}~\bibnamefont {Venugopalan}},\
  }\href@noop {} \Eprint
  {http://arxiv.org/abs/1601.07342} {arXiv:1601.07342 [hep-ph]} \BibitemShut
  {NoStop}%
\bibitem [{\citenamefont {Mueller}\ \emph {et~al.}()\citenamefont {Mueller},
  \citenamefont {Hebenstreit},\ and\ \citenamefont {Berges}}]{Mueller:2016}%
  \BibitemOpen
  \bibfield  {author} {\bibinfo {author} {\bibfnamefont {N.}~\bibnamefont
  {Mueller}}, \bibinfo {author} {\bibfnamefont {F.}~\bibnamefont
  {Hebenstreit}}, \ and\ \bibinfo {author} {\bibfnamefont {J.}~\bibnamefont
  {Berges}},\ }\href@noop {} {\bibinfo  {journal} {(to be published)}\
  }\BibitemShut {NoStop}%
\bibitem [{\citenamefont {Adler}(1969)}]{Adler:1969gk}%
  \BibitemOpen
\bibfield  {journal} {  }\bibfield  {author} {\bibinfo {author} {\bibfnamefont
  {S.~L.}\ \bibnamefont {Adler}},\ }\href {\doibase 10.1103/PhysRev.177.2426}
  {\bibfield  {journal} {\bibinfo  {journal} {Phys. Rev.}\ }\textbf {\bibinfo
  {volume} {177}},\ \bibinfo {pages} {2426} (\bibinfo {year}
  {1969})}\BibitemShut {NoStop}%
\bibitem [{\citenamefont {Bell}\ and\ \citenamefont
  {Jackiw}(1969)}]{Bell:1969ts}%
  \BibitemOpen
  \bibfield  {author} {\bibinfo {author} {\bibfnamefont {J.~S.}\ \bibnamefont
  {Bell}}\ and\ \bibinfo {author} {\bibfnamefont {R.}~\bibnamefont {Jackiw}},\
  }\href {\doibase 10.1007/BF02823296} {\bibfield  {journal} {\bibinfo
  {journal} {Nuovo Cim.}\ }\textbf {\bibinfo {volume} {A60}},\ \bibinfo {pages}
  {47} (\bibinfo {year} {1969})}\BibitemShut {NoStop}%
\bibitem [{\citenamefont {Baseyan}\ \emph {et~al.}(1979)\citenamefont
  {Baseyan}, \citenamefont {Matinyan},\ and\ \citenamefont
  {Savvidi}}]{baseyan1979nonlinear}%
  \BibitemOpen
  \bibfield  {author} {\bibinfo {author} {\bibfnamefont {G.}~\bibnamefont
  {Baseyan}}, \bibinfo {author} {\bibfnamefont {S.}~\bibnamefont {Matinyan}}, \
  and\ \bibinfo {author} {\bibfnamefont {G.}~\bibnamefont {Savvidi}},\
  }\href@noop {} {\bibfield  {journal} {\bibinfo  {journal} {Pisma Zh. Eksp.
  Teor. Fiz.}\ }\textbf {\bibinfo {volume} {29}},\ \bibinfo {pages} {641}
  (\bibinfo {year} {1979})},\ \bibinfo {note} {[JETP Lett. \textbf{29}, 587
  (1979)]}\BibitemShut {NoStop}%
\bibitem [{\citenamefont {Matinyan}\ \emph {et~al.}(1981)\citenamefont
  {Matinyan}, \citenamefont {Savvidi},\ and\ \citenamefont
  {Ter-Arutyunyan-Savvidi}}]{matinyan1981classical}%
  \BibitemOpen
  \bibfield  {author} {\bibinfo {author} {\bibfnamefont {S.}~\bibnamefont
  {Matinyan}}, \bibinfo {author} {\bibfnamefont {G.}~\bibnamefont {Savvidi}}, \
  and\ \bibinfo {author} {\bibfnamefont {N.}~\bibnamefont
  {Ter-Arutyunyan-Savvidi}},\ }\href@noop {} {\bibfield  {journal} {\bibinfo
  {journal} {Zh. Eksp. Teor. Fiz.}\ }\textbf {\bibinfo {volume} {80}},\
  \bibinfo {pages} {830} (\bibinfo {year} {1981})},\ \bibinfo {note} {[Sov.
  Phys. JETP. \textbf{53}, 421 (1981)]}\BibitemShut {NoStop}%
\bibitem [{\citenamefont {Bir{\'o}}\ \emph {et~al.}()\citenamefont {Bir{\'o}},
  \citenamefont {Matinyan},\ and\ \citenamefont {M{\"u}ller}}]{biro1995chaos}%
  \BibitemOpen
  \bibfield  {author} {\bibinfo {author} {\bibfnamefont {T.~S.}\ \bibnamefont
  {Bir{\'o}}}, \bibinfo {author} {\bibfnamefont {S.~G.}\ \bibnamefont
  {Matinyan}}, \ and\ \bibinfo {author} {\bibfnamefont {B.}~\bibnamefont
  {M{\"u}ller}},\ }\href@noop {} {\bibinfo  {journal} {\textit{Chaos and Gauge Field
  Theory} (World Scientific, Singapore, 1994).}\ }\BibitemShut {NoStop}%
\bibitem [{\citenamefont {Aarts}\ and\ \citenamefont
  {Smit}(1999)}]{Aarts:1998td}%
  \BibitemOpen
\bibfield  {journal} {  }\bibfield  {author} {\bibinfo {author} {\bibfnamefont
  {G.}~\bibnamefont {Aarts}}\ and\ \bibinfo {author} {\bibfnamefont
  {J.}~\bibnamefont {Smit}},\ }\href {\doibase 10.1016/S0550-3213(99)00320-X}
  {\bibfield  {journal} {\bibinfo  {journal} {Nucl. Phys.}\ }\textbf {\bibinfo
  {volume} {B555}},\ \bibinfo {pages} {355} (\bibinfo {year}
  {1999})}\BibitemShut {NoStop}%
\bibitem [{\citenamefont {Borsanyi}\ and\ \citenamefont
  {Hindmarsh}(2009)}]{Borsanyi:2008eu}%
  \BibitemOpen
  \bibfield  {author} {\bibinfo {author} {\bibfnamefont {S.}~\bibnamefont
  {Borsanyi}}\ and\ \bibinfo {author} {\bibfnamefont {M.}~\bibnamefont
  {Hindmarsh}},\ }\href {\doibase 10.1103/PhysRevD.79.065010} {\bibfield
  {journal} {\bibinfo  {journal} {Phys. Rev.}\ }\textbf {\bibinfo {volume}
  {D79}},\ \bibinfo {pages} {065010} (\bibinfo {year} {2009})}\BibitemShut
  {NoStop}%
\bibitem [{\citenamefont {Berges}\ \emph {et~al.}(2011)\citenamefont {Berges},
  \citenamefont {Gelfand},\ and\ \citenamefont {Pruschke}}]{Berges:2010zv}%
  \BibitemOpen
  \bibfield  {author} {\bibinfo {author} {\bibfnamefont {J.}~\bibnamefont
  {Berges}}, \bibinfo {author} {\bibfnamefont {D.}~\bibnamefont {Gelfand}}, \
  and\ \bibinfo {author} {\bibfnamefont {J.}~\bibnamefont {Pruschke}},\ }\href
  {\doibase 10.1103/PhysRevLett.107.061301} {\bibfield  {journal} {\bibinfo
  {journal} {Phys. Rev. Lett.}\ }\textbf {\bibinfo {volume} {107}},\ \bibinfo
  {pages} {061301} (\bibinfo {year} {2011})}\BibitemShut {NoStop}%
\bibitem [{\citenamefont {Saffin}\ and\ \citenamefont
  {Tranberg}(2011)}]{Saffin:2011kc}%
  \BibitemOpen
  \bibfield  {author} {\bibinfo {author} {\bibfnamefont {P.~M.}\ \bibnamefont
  {Saffin}}\ and\ \bibinfo {author} {\bibfnamefont {A.}~\bibnamefont
  {Tranberg}},\ }\href {\doibase 10.1007/JHEP07(2011)066} {\bibfield  {journal}
  {\bibinfo  {journal} {JHEP}\ }\textbf {\bibinfo {volume} {07}},\ \bibinfo
  {pages} {066} (\bibinfo {year} {2011})}\BibitemShut {NoStop}%
\bibitem [{\citenamefont {Hebenstreit}\ \emph {et~al.}(2013)\citenamefont
  {Hebenstreit}, \citenamefont {Berges},\ and\ \citenamefont
  {Gelfand}}]{Hebenstreit:2013qxa}%
  \BibitemOpen
  \bibfield  {author} {\bibinfo {author} {\bibfnamefont {F.}~\bibnamefont
  {Hebenstreit}}, \bibinfo {author} {\bibfnamefont {J.}~\bibnamefont {Berges}},
  \ and\ \bibinfo {author} {\bibfnamefont {D.}~\bibnamefont {Gelfand}},\ }\href
  {\doibase 10.1103/PhysRevD.87.105006} {\bibfield  {journal} {\bibinfo
  {journal} {Phys. Rev.}\ }\textbf {\bibinfo {volume} {D87}},\ \bibinfo {pages}
  {105006} (\bibinfo {year} {2013})}\BibitemShut {NoStop}%
\bibitem [{\citenamefont {Kasper}\ \emph {et~al.}(2014)\citenamefont {Kasper},
  \citenamefont {Hebenstreit},\ and\ \citenamefont {Berges}}]{Kasper:2014uaa}%
  \BibitemOpen
  \bibfield  {author} {\bibinfo {author} {\bibfnamefont {V.}~\bibnamefont
  {Kasper}}, \bibinfo {author} {\bibfnamefont {F.}~\bibnamefont {Hebenstreit}},
  \ and\ \bibinfo {author} {\bibfnamefont {J.}~\bibnamefont {Berges}},\ }\href
  {\doibase 10.1103/PhysRevD.90.025016} {\bibfield  {journal} {\bibinfo
  {journal} {Phys. Rev.}\ }\textbf {\bibinfo {volume} {D90}},\ \bibinfo {pages}
  {025016} (\bibinfo {year} {2014})}\BibitemShut {NoStop}%
\bibitem [{\citenamefont {Tanji}(2015)}]{Tanji:2015ata}%
  \BibitemOpen
  \bibfield  {author} {\bibinfo {author} {\bibfnamefont {N.}~\bibnamefont
  {Tanji}},\ }\href {\doibase 10.1103/PhysRevD.92.125012} {\bibfield  {journal}
  {\bibinfo  {journal} {Phys. Rev.}\ }\textbf {\bibinfo {volume} {D92}},\
  \bibinfo {pages} {125012} (\bibinfo {year} {2015})}\BibitemShut {NoStop}%
\bibitem [{\citenamefont {Gelis}\ and\ \citenamefont
  {Tanji}(2016)}]{Gelis:2015kya}%
  \BibitemOpen
  \bibfield  {author} {\bibinfo {author} {\bibfnamefont {F.}~\bibnamefont
  {Gelis}}\ and\ \bibinfo {author} {\bibfnamefont {N.}~\bibnamefont {Tanji}},\
  }\href {\doibase 10.1016/j.ppnp.2015.11.001} {\bibfield  {journal} {\bibinfo
  {journal} {Prog. Part. Nucl. Phys.}\ }\textbf {\bibinfo {volume} {87}},\
  \bibinfo {pages} {1} (\bibinfo {year} {2016})}\BibitemShut {NoStop}%
\bibitem [{\citenamefont {Gelfand}\ \emph {et~al.}(2016)\citenamefont
  {Gelfand}, \citenamefont {Hebenstreit},\ and\ \citenamefont
  {Berges}}]{Gelfand:2016prm}%
  \BibitemOpen
  \bibfield  {author} {\bibinfo {author} {\bibfnamefont {D.}~\bibnamefont
  {Gelfand}}, \bibinfo {author} {\bibfnamefont {F.}~\bibnamefont
  {Hebenstreit}}, \ and\ \bibinfo {author} {\bibfnamefont {J.}~\bibnamefont
  {Berges}},\ }\href {\doibase 10.1103/PhysRevD.93.085001} {\bibfield
  {journal} {\bibinfo  {journal} {Phys. Rev.}\ }\textbf {\bibinfo {volume}
  {D93}},\ \bibinfo {pages} {085001} (\bibinfo {year} {2016})}\BibitemShut
  {NoStop}%
\bibitem [{\citenamefont {Karsten}\ and\ \citenamefont
  {Smit}(1981)}]{Karsten:1980wd}%
  \BibitemOpen
  \bibfield  {author} {\bibinfo {author} {\bibfnamefont {L.~H.}\ \bibnamefont
  {Karsten}}\ and\ \bibinfo {author} {\bibfnamefont {J.}~\bibnamefont {Smit}},\
  }\href {\doibase 10.1016/0550-3213(81)90549-6} {\bibfield  {journal}
  {\bibinfo  {journal} {Nucl.Phys.}\ }\textbf {\bibinfo {volume} {B183}},\
  \bibinfo {pages} {103} (\bibinfo {year} {1981})}\BibitemShut {NoStop}%
\bibitem [{\citenamefont {Nielsen}\ and\ \citenamefont
  {Ninomiya}(1981)}]{Nielsen:1980rz}%
  \BibitemOpen
  \bibfield  {author} {\bibinfo {author} {\bibfnamefont {H.~B.}\ \bibnamefont
  {Nielsen}}\ and\ \bibinfo {author} {\bibfnamefont {M.}~\bibnamefont
  {Ninomiya}},\ }\href {\doibase 10.1016/0550-3213(81)90361-8} {\bibfield
  {journal} {\bibinfo  {journal} {Nucl.Phys.}\ }\textbf {\bibinfo {volume}
  {B185}},\ \bibinfo {pages} {20} (\bibinfo {year} {1981})}\BibitemShut
  {NoStop}%
\bibitem [{\citenamefont {Friedan}(1982)}]{Friedan:1982nk}%
  \BibitemOpen
  \bibfield  {author} {\bibinfo {author} {\bibfnamefont {D.}~\bibnamefont
  {Friedan}},\ }\href {\doibase 10.1007/BF01403500} {\bibfield  {journal}
  {\bibinfo  {journal} {Commun. Math. Phys.}\ }\textbf {\bibinfo {volume}
  {85}},\ \bibinfo {pages} {481} (\bibinfo {year} {1982})}\BibitemShut
  {NoStop}%
\bibitem [{\citenamefont {Kharzeev}\ \emph {et~al.}(2002)\citenamefont
  {Kharzeev}, \citenamefont {Krasnitz},\ and\ \citenamefont
  {Venugopalan}}]{Kharzeev:2001ev}%
  \BibitemOpen
  \bibfield  {author} {\bibinfo {author} {\bibfnamefont {D.}~\bibnamefont
  {Kharzeev}}, \bibinfo {author} {\bibfnamefont {A.}~\bibnamefont {Krasnitz}},
  \ and\ \bibinfo {author} {\bibfnamefont {R.}~\bibnamefont {Venugopalan}},\
  }\href {\doibase 10.1016/S0370-2693(02)02630-8} {\bibfield  {journal}
  {\bibinfo  {journal} {Phys. Lett.}\ }\textbf {\bibinfo {volume} {B545}},\
  \bibinfo {pages} {298} (\bibinfo {year} {2002})}\BibitemShut {NoStop}%
\bibitem [{\citenamefont {Berges}\ \emph {et~al.}(2012)\citenamefont {Berges},
  \citenamefont {Scheffler}, \citenamefont {Schlichting},\ and\ \citenamefont
  {Sexty}}]{Berges:2011sb}%
  \BibitemOpen
  \bibfield  {author} {\bibinfo {author} {\bibfnamefont {J.}~\bibnamefont
  {Berges}}, \bibinfo {author} {\bibfnamefont {S.}~\bibnamefont {Scheffler}},
  \bibinfo {author} {\bibfnamefont {S.}~\bibnamefont {Schlichting}}, \ and\
  \bibinfo {author} {\bibfnamefont {D.}~\bibnamefont {Sexty}},\ }\href
  {\doibase 10.1103/PhysRevD.85.034507} {\bibfield  {journal} {\bibinfo
  {journal} {Phys. Rev.}\ }\textbf {\bibinfo {volume} {D85}},\ \bibinfo {pages}
  {034507} (\bibinfo {year} {2012})}\BibitemShut {NoStop}%
\bibitem [{\citenamefont {Tudron}(1980)}]{Tudron:1980gq}%
  \BibitemOpen
  \bibfield  {author} {\bibinfo {author} {\bibfnamefont {T.~N.}\ \bibnamefont
  {Tudron}},\ }\href {\doibase 10.1103/PhysRevD.22.2566} {\bibfield  {journal}
  {\bibinfo  {journal} {Phys. Rev.}\ }\textbf {\bibinfo {volume} {D22}},\
  \bibinfo {pages} {2566} (\bibinfo {year} {1980})}\BibitemShut {NoStop}%
\bibitem [{\citenamefont {Abramowitz}\ and\ \citenamefont
  {Stegun}()}]{abramowitz1965handbook}%
  \BibitemOpen
  \bibfield  {author} {\bibinfo {author} {\bibfnamefont {M.}~\bibnamefont
  {Abramowitz}}\ and\ \bibinfo {author} {\bibfnamefont {I.~A.}\ \bibnamefont
  {Stegun}},\ }\href@noop {} {\bibinfo  {journal} {\textit{Handbook of Mathematical
  Functions} (Dover Publications, New York, 1965).}\ }\BibitemShut {NoStop}%
\bibitem [{\citenamefont {Nielsen}\ and\ \citenamefont
  {Olesen}(1978)}]{Nielsen:1978rm}%
  \BibitemOpen
\bibfield  {journal} {  }\bibfield  {author} {\bibinfo {author} {\bibfnamefont
  {N.~K.}\ \bibnamefont {Nielsen}}\ and\ \bibinfo {author} {\bibfnamefont
  {P.}~\bibnamefont {Olesen}},\ }\href {\doibase 10.1016/0550-3213(78)90377-2}
  {\bibfield  {journal} {\bibinfo  {journal} {Nucl. Phys.}\ }\textbf {\bibinfo
  {volume} {B144}},\ \bibinfo {pages} {376} (\bibinfo {year}
  {1978})}\BibitemShut {NoStop}%
\bibitem [{\citenamefont {Chang}\ and\ \citenamefont
  {Weiss}(1979)}]{Chang:1979tg}%
  \BibitemOpen
  \bibfield  {author} {\bibinfo {author} {\bibfnamefont {S.-J.}\ \bibnamefont
  {Chang}}\ and\ \bibinfo {author} {\bibfnamefont {N.}~\bibnamefont {Weiss}},\
  }\href {\doibase 10.1103/PhysRevD.20.869} {\bibfield  {journal} {\bibinfo
  {journal} {Phys. Rev.}\ }\textbf {\bibinfo {volume} {D20}},\ \bibinfo {pages}
  {869} (\bibinfo {year} {1979})}\BibitemShut {NoStop}%
\bibitem [{\citenamefont {Romatschke}\ and\ \citenamefont
  {Venugopalan}(2006)}]{Romatschke:2006nk}%
  \BibitemOpen
  \bibfield  {author} {\bibinfo {author} {\bibfnamefont {P.}~\bibnamefont
  {Romatschke}}\ and\ \bibinfo {author} {\bibfnamefont {R.}~\bibnamefont
  {Venugopalan}},\ }\href {\doibase 10.1103/PhysRevD.74.045011} {\bibfield
  {journal} {\bibinfo  {journal} {Phys. Rev.}\ }\textbf {\bibinfo {volume}
  {D74}},\ \bibinfo {pages} {045011} (\bibinfo {year} {2006})}\BibitemShut
  {NoStop}%
\bibitem [{\citenamefont {Berges}\ \emph {et~al.}(2008)\citenamefont {Berges},
  \citenamefont {Scheffler},\ and\ \citenamefont {Sexty}}]{Berges:2007re}%
  \BibitemOpen
  \bibfield  {author} {\bibinfo {author} {\bibfnamefont {J.}~\bibnamefont
  {Berges}}, \bibinfo {author} {\bibfnamefont {S.}~\bibnamefont {Scheffler}}, \
  and\ \bibinfo {author} {\bibfnamefont {D.}~\bibnamefont {Sexty}},\ }\href
  {\doibase 10.1103/PhysRevD.77.034504} {\bibfield  {journal} {\bibinfo
  {journal} {Phys. Rev.}\ }\textbf {\bibinfo {volume} {D77}},\ \bibinfo {pages}
  {034504} (\bibinfo {year} {2008})}\BibitemShut {NoStop}%
\bibitem [{\citenamefont {Kurkela}\ and\ \citenamefont
  {Moore}(2012)}]{Kurkela:2012hp}%
  \BibitemOpen
  \bibfield  {author} {\bibinfo {author} {\bibfnamefont {A.}~\bibnamefont
  {Kurkela}}\ and\ \bibinfo {author} {\bibfnamefont {G.~D.}\ \bibnamefont
  {Moore}},\ }\href {\doibase 10.1103/PhysRevD.86.056008} {\bibfield  {journal}
  {\bibinfo  {journal} {Phys. Rev.}\ }\textbf {\bibinfo {volume} {D86}},\
  \bibinfo {pages} {056008} (\bibinfo {year} {2012})}\BibitemShut {NoStop}%
\bibitem [{\citenamefont {Berges}\ \emph {et~al.}(2014)\citenamefont {Berges},
  \citenamefont {Boguslavski}, \citenamefont {Schlichting},\ and\ \citenamefont
  {Venugopalan}}]{Berges:2013eia}%
  \BibitemOpen
  \bibfield  {author} {\bibinfo {author} {\bibfnamefont {J.}~\bibnamefont
  {Berges}}, \bibinfo {author} {\bibfnamefont {K.}~\bibnamefont {Boguslavski}},
  \bibinfo {author} {\bibfnamefont {S.}~\bibnamefont {Schlichting}}, \ and\
  \bibinfo {author} {\bibfnamefont {R.}~\bibnamefont {Venugopalan}},\ }\href
  {\doibase 10.1103/PhysRevD.89.074011} {\bibfield  {journal} {\bibinfo
  {journal} {Phys. Rev.}\ }\textbf {\bibinfo {volume} {D89}},\ \bibinfo {pages}
  {074011} (\bibinfo {year} {2014})}\BibitemShut {NoStop}%
\bibitem [{\citenamefont {Epelbaum}\ and\ \citenamefont
  {Gelis}(2013)}]{Gelis:2013rba}%
  \BibitemOpen
  \bibfield  {author} {\bibinfo {author} {\bibfnamefont {T.}~\bibnamefont
  {Epelbaum}}\ and\ \bibinfo {author} {\bibfnamefont {F.}~\bibnamefont
  {Gelis}},\ }\href {\doibase 10.1103/PhysRevLett.111.232301} {\bibfield
  {journal} {\bibinfo  {journal} {Phys. Rev. Lett.}\ }\textbf {\bibinfo
  {volume} {111}},\ \bibinfo {pages} {232301} (\bibinfo {year}
  {2013})}\BibitemShut {NoStop}%
\bibitem [{\citenamefont {Moore}(1996{\natexlab{a}})}]{Moore:1996qs}%
  \BibitemOpen
  \bibfield  {author} {\bibinfo {author} {\bibfnamefont {G.~D.}\ \bibnamefont
  {Moore}},\ }\href {\doibase 10.1016/S0550-3213(96)00445-2} {\bibfield
  {journal} {\bibinfo  {journal} {Nucl. Phys.}\ }\textbf {\bibinfo {volume}
  {B480}},\ \bibinfo {pages} {657} (\bibinfo {year}
  {1996}{\natexlab{a}})}\BibitemShut {NoStop}%
\bibitem [{\citenamefont {Moore}(1996{\natexlab{b}})}]{Moore:1996wn}%
  \BibitemOpen
  \bibfield  {author} {\bibinfo {author} {\bibfnamefont {G.~D.}\ \bibnamefont
  {Moore}},\ }\href {\doibase 10.1016/S0550-3213(96)00497-X} {\bibfield
  {journal} {\bibinfo  {journal} {Nucl. Phys.}\ }\textbf {\bibinfo {volume}
  {B480}},\ \bibinfo {pages} {689} (\bibinfo {year}
  {1996}{\natexlab{b}})}\BibitemShut {NoStop}%
\bibitem [{\citenamefont {Buividovich}\ and\ \citenamefont
  {Ulybyshev}(2015)}]{Buividovich:2015jfa}%
  \BibitemOpen
  \bibfield  {author} {\bibinfo {author} {\bibfnamefont {P.~V.}\ \bibnamefont
  {Buividovich}}\ and\ \bibinfo {author} {\bibfnamefont {M.~V.}\ \bibnamefont
  {Ulybyshev}},\ }\href@noop {} \Eprint
  {http://arxiv.org/abs/1509.02076} {arXiv:1509.02076 [hep-th]} \BibitemShut
  {NoStop}%
\bibitem [{\citenamefont {Rothe}\ and\ \citenamefont
  {Sadooghi}(1998)}]{Rothe:1998ba}%
  \BibitemOpen
  \bibfield  {author} {\bibinfo {author} {\bibfnamefont {H.~J.}\ \bibnamefont
  {Rothe}}\ and\ \bibinfo {author} {\bibfnamefont {N.}~\bibnamefont
  {Sadooghi}},\ }\href {\doibase 10.1103/PhysRevD.58.074502} {\bibfield
  {journal} {\bibinfo  {journal} {Phys.Rev.}\ }\textbf {\bibinfo {volume}
  {D58}},\ \bibinfo {pages} {074502} (\bibinfo {year} {1998})}\BibitemShut
  {NoStop}%
\bibitem [{\citenamefont {Reisz}\ and\ \citenamefont
  {Rothe}(1999)}]{Reisz:1999cma}%
  \BibitemOpen
  \bibfield  {author} {\bibinfo {author} {\bibfnamefont {T.}~\bibnamefont
  {Reisz}}\ and\ \bibinfo {author} {\bibfnamefont {H.}~\bibnamefont {Rothe}},\
  }\href {\doibase 10.1016/S0370-2693(99)00491-8} {\bibfield  {journal}
  {\bibinfo  {journal} {Phys.Lett.}\ }\textbf {\bibinfo {volume} {B455}},\
  \bibinfo {pages} {246} (\bibinfo {year} {1999})}\BibitemShut {NoStop}%
\bibitem [{\citenamefont {Akamatsu}\ and\ \citenamefont
  {Yamamoto}(2013)}]{Akamatsu:2013pjd}%
  \BibitemOpen
  \bibfield  {author} {\bibinfo {author} {\bibfnamefont {Y.}~\bibnamefont
  {Akamatsu}}\ and\ \bibinfo {author} {\bibfnamefont {N.}~\bibnamefont
  {Yamamoto}},\ }\href {\doibase 10.1103/PhysRevLett.111.052002} {\bibfield
  {journal} {\bibinfo  {journal} {Phys. Rev. Lett.}\ }\textbf {\bibinfo
  {volume} {111}},\ \bibinfo {pages} {052002} (\bibinfo {year}
  {2013})}\BibitemShut {NoStop}%
\bibitem [{\citenamefont {Akamatsu}\ \emph {et~al.}(2016)\citenamefont
  {Akamatsu}, \citenamefont {Rothkopf},\ and\ \citenamefont
  {Yamamoto}}]{Akamatsu:2015kau}%
  \BibitemOpen
  \bibfield  {author} {\bibinfo {author} {\bibfnamefont {Y.}~\bibnamefont
  {Akamatsu}}, \bibinfo {author} {\bibfnamefont {A.}~\bibnamefont {Rothkopf}},
  \ and\ \bibinfo {author} {\bibfnamefont {N.}~\bibnamefont {Yamamoto}},\
  }\href@noop {} {\bibfield  {journal} {\bibinfo  {journal} {JHEP}\ }\textbf
  {\bibinfo {volume} {03}},\ \bibinfo {pages} {210} (\bibinfo {year}
  {2016})}\BibitemShut {NoStop}%
\bibitem [{\citenamefont {Fukushima}(2015)}]{Fukushima:2015tza}%
  \BibitemOpen
  \bibfield  {author} {\bibinfo {author} {\bibfnamefont {K.}~\bibnamefont
  {Fukushima}},\ }\href {\doibase 10.1103/PhysRevD.92.054009} {\bibfield
  {journal} {\bibinfo  {journal} {Phys. Rev.}\ }\textbf {\bibinfo {volume}
  {D92}},\ \bibinfo {pages} {054009} (\bibinfo {year} {2015})}\BibitemShut
  {NoStop}%
\bibitem [{\citenamefont {Klinkhamer}\ and\ \citenamefont
  {Rupp}(2003)}]{Klinkhamer:2003hz}%
  \BibitemOpen
  \bibfield  {author} {\bibinfo {author} {\bibfnamefont {F.~R.}\ \bibnamefont
  {Klinkhamer}}\ and\ \bibinfo {author} {\bibfnamefont {C.}~\bibnamefont
  {Rupp}},\ }\href {\doibase 10.1063/1.1590420} {\bibfield  {journal} {\bibinfo
   {journal} {J. Math. Phys.}\ }\textbf {\bibinfo {volume} {44}},\ \bibinfo
  {pages} {3619} (\bibinfo {year} {2003})}\BibitemShut {NoStop}%
\end{thebibliography}

%

\end{document}